\documentclass[twocolumn]{aastex7}

\usepackage{amsmath}

\usepackage{xcolor}

\shorttitle{Modeling of X-ray Foregrounds, Backgrounds and Faint Sources}
\shortauthors{Mantz et al.}

\graphicspath{{./}{figures/}}

\begin{document}

\title{Ruminations Upon the Modeling of X-ray Foregrounds, Backgrounds and Faint Sources}

\author[0000-0002-8031-1217]{Adam B. Mantz}
\affiliation{Kavli Institute for Particle Astrophysics and Cosmology, Stanford University, 452 Lomita Mall, Stanford, CA 94305, USA}
\email{amantz@stanford.edu}
\correspondingauthor{amantz@stanford.edu}

\author[0000-0001-7179-6198]{Anthony M. Flores}
\affiliation{Kavli Institute for Particle Astrophysics and Cosmology, Stanford University, 452 Lomita Mall, Stanford, CA 94305, USA}
\affiliation{Department of Physics, Stanford University, 382 Via Pueblo Mall, Stanford, CA 94305, USA}
\affiliation{SLAC National Accelerator Laboratory, 2575 Sand Hill Road, Menlo Park, CA  94025, USA}
\affiliation{Department of Physics and Astronomy, Rutgers University, 136 Frelinghuysen Road, Piscataway, NJ 08854, USA}
\email{aflores7@stanford.edu}

\author[0000-0003-3521-3631]{Taweewat Somboonpanyakul}
\affiliation{Department of Physics, Faculty of Science, Chulalongkorn University, 254 Phyathai Road, Patumwan, Bangkok 10330, Thailand}
\email{taweewat.s@chula.ac.th}

\author[0000-0003-0667-5941]{Steven W. Allen}
\affiliation{Kavli Institute for Particle Astrophysics and Cosmology, Stanford University, 452 Lomita Mall, Stanford, CA 94305, USA}
\affiliation{Department of Physics, Stanford University, 382 Via Pueblo Mall, Stanford, CA 94305, USA}
\affiliation{SLAC National Accelerator Laboratory, 2575 Sand Hill Road, Menlo Park, CA  94025, USA}
\email{swa@stanford.edu}

\author[0000-0003-2985-9962]{R. Glenn Morris}
\affiliation{Kavli Institute for Particle Astrophysics and Cosmology, Stanford University, 452 Lomita Mall, Stanford, CA 94305, USA}
\affiliation{SLAC National Accelerator Laboratory, 2575 Sand Hill Road, Menlo Park, CA  94025, USA}
\email{rgm@stanford.edu}

\author[0009-0001-9176-8861]{Abigail Y. Pan}
\affiliation{Kavli Institute for Particle Astrophysics and Cosmology, Stanford University, 452 Lomita Mall, Stanford, CA 94305, USA}
\affiliation{Department of Physics, Stanford University, 382 Via Pueblo Mall, Stanford, CA 94305, USA}
\email{apan5@stanford.edu}

\author[0000-0002-2776-978X]{Haley R. Stueber}
\affiliation{Kavli Institute for Particle Astrophysics and Cosmology, Stanford University, 452 Lomita Mall, Stanford, CA 94305, USA}
\affiliation{Department of Physics, Stanford University, 382 Via Pueblo Mall, Stanford, CA 94305, USA}
\email{hstueber@stanford.edu}


\begin{abstract}
  With the goal of extracting as much information as possible from Chandra and XMM-Newton observations of faint, diffuse sources such as galaxy clusters, as well as those of future X-ray telescopes, we present a strategy for forward modeling all the foreground and background signals present in these data.
  This work leverages widespread efforts to understand the soft X-ray emission from the Galaxy, as well as the cosmic X-ray background and instrument-specific, particle-induced backgrounds.
  Statistically, a forward model of the foregrounds and backgrounds is preferable to alternatives because it requires no binning of the data, and allows straightforward marginalization over systematic uncertainties.
  We apply these methods to several galaxy clusters at intermediate-to-high redshifts, spanning a range of masses and morphologies, using Chandra and/or XMM-Newton data.
  Our results suggest a modest improvement even for relatively bright clusters at these redshifts, and more substantial advantages in the high-redshift, low-surface-brightness regime.
  We also discuss and provide a simple correction for a time-dependent miscalibration of the Chandra ACIS detectors identified in archival galaxy cluster data.
\end{abstract}


\section{Introduction} \label{sec:intro}



Regardless of their purpose, astronomical observations in the soft X-ray band (order 0.1--10\,keV) are affected by multiple, ubiquitous signals, including astrophysical X-rays originating within the Solar system and the Galaxy, extragalactic X-ray sources, and events caused by energetic particles.
The former component, hereafter the soft foreground (SFG), is thought to consist primarily of thermal emission from the local hot bubble and the Galactic halo (e.g.\ \citealt{Snowden1998ApJ...493..715S, Kuntz2000ApJ...543..195K}), along with a time-variable contribution from solar wind charge exchange (SWCX; e.g.\ \citealt{Snowden0404354, Koutroumpa0805.3212, Ueda2209.01698}).
The bulk of the extragalactic background (hereafter the cosmic X-ray background, CXB) is due to active galactic nuclei (AGN), the brightest subset of which can be detected and masked in observations with sufficiently high spatial resolution \citep{Hornschemeier0004260, Mushotzky0002313, Brandt0108404, Giacconi0007240, Rosati2002ApJ...566..667R}.
Particle-induced events depend on the detector and observatory involved, and can include secondary particles and true X-rays produced in showers when cosmic rays interact with material in or near the detector.
The rate of such events is time variable, with a ``quiescent'' particle background (QPB) rate closely following the Solar cycle, as well as unpredictable periods of higher background, potentially by orders of magnitude.
Standard practice when studying extended X-ray sources is to use only times when the particle background is consistent with the quiescent level, and to mask those CXB sources that are bright enough to individually detect in a given observation.
The remaining foreground and background signals\footnote{For brevity, we will henceforth use ``background'' to refer to both foregrounds and backgrounds unless the distinction is important.} can nevertheless be comparable to, if not brighter than, the source of interest in some circumstances, such as the intracluster medium (ICM) at large radii or high redshifts, or the circumgalactic medium of other galaxies.

The application of immediate interest in this work is studies of the ICM in galaxy clusters at intermediate-to-high redshifts with Chandra advanced CCD imaging spectrometer (ACIS) or XMM-Newton (hereafter XMM) European photon imaging camera (EPIC) data.
In this area, typical observations yield of order $10^3$--$10^4$ source counts in total, with a comparable or greater number of background events in the region of interest, after standard filtering to eliminate cosmic ray primaries and times of non-quiescent background.
Divided among thousands of pixels and/or spectral channels, the Poisson nature of the data cannot be ignored.
This feature, combined with the fact that each of the background components above can vary either spatially or temporally (or both), makes the process of estimating and accounting for the background challenging.

The simplest and most traditional approach is to obtain a spectrum from the observation being analyzed, using part of the detector free of emission from the source of interest (if possible).
By construction, this background estimate is temporally appropriate, being concurrent with the source spectrum; however, it is necessarily produced from a different region of the detector.
A common alternative is to instead estimate the background using the same part of the detector but a different observation, in which the region in question contains no (non-background) sources.
Typically these estimates are rescaled based on the count rate at energies higher than the optics function (i.e., where all recorded events are due to particles) in order to account for long-term variability in the normalization of the QPB.
Key remaining considerations are then how well the spectral shape of the QPB can be expected to match between the two observations, as well as whether the background observation also includes the astrophysical X-ray backgrounds.
If so, additional modeling may be required to account for differences in the local SFG, or in the depth to which CXB sources can be detected and removed, or simply because these astrophysical components do not vary temporally with the QPB yet have been renormalized along with it.
Furthermore, the utility of a separate background observation is hindered if it is not significantly longer than the science observation, such that its statistical variance is smaller.

Direct application of such empirical background estimates is also problematic from a statistical perspective.
The use of a Gaussian likelihood function to fit fundamentally Poisson data biases results, even when binning the data to hundreds of counts or more \citep{Cash1979ApJ...228..939, Wachter1979ApJ...230..274W, Nousek1989ApJ...342.1207N, Leccardi0705.4199, Humphrey0811.2796, Mantz1706.01476, Bonamente1912.05444}.
Use of a Poisson likelihood (equivalently, the Cash fitting statistic; \citealt{Cash1979ApJ...228..939}) ameliorates this issue but, unlike a Gaussian likelihood, it does not provide a simple way to directly subtract a background estimated with a given (also Gaussian) uncertainty.
A common solution in spectroscopy is to use the modified Cash statistic which approximately marginalizes over the background signal in each channel, as constrained by a separate, empirical and similarly Poisson measurement of the background.\footnote{\url{https://heasarc.gsfc.nasa.gov/docs/software/xspec/manual/node336.html}}
This approach produces unbiased measurements in practice \citep{Humphrey0811.2796, Mantz1706.01476, Bonamente1912.05444}, though it still requires data to be binned to $\geq1$ count to do so.\footnote{We are not aware of a refereed citation for this fact, but it can be straightforwardly verified by simulation.}
While this may not seem like a great imposition, it can be a limitation for imaging-spectroscopic applications, requiring one to compromise either spatial or spectral resolution unnecessarily.
Binning aside, the notions that we have so little information about the background (i.e., no more than is present in the background region of a particular observation) and that the background in every channel is independent of every other strain credulity (see also discussions by \citealt{van-Dyk0008170, Broos1003.2397}).

\begin{figure}
    \centering
    \includegraphics{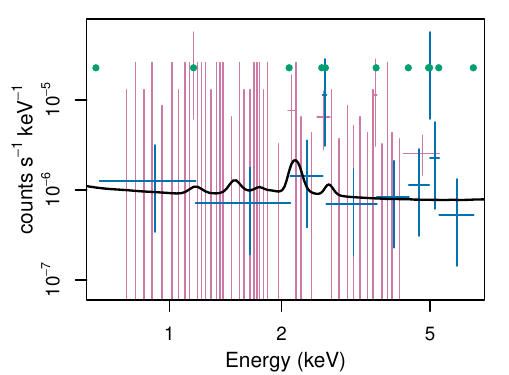}
    \caption{%
      Visualization of a particular Chandra blank-sky spectrum for a $2''$ radius circle, compared with the corresponding generative model.
      Green points (corresponding to individual events) show the unbinned spectrum.
      Blue crosses show the spectrum grouped to a minimum of 1 count per bin using the ``group min 1'' command to the {\sc grppha} tool.
      Pink crosses show the same spectrum binned according to the same grouping procedure applied to the corresponding science spectrum.
      Error bars on these points show the 68.3 percent credible intervals on the spectrum in each bin, assuming uniform and independent priors over the non-negative real line.
      The black curve shows the model produced by our modification of the \citet{Suzuki2108.11234} procedure (see Section~\ref{sec:chqpb}).
    }
    \label{fig:sparse_bg}
\end{figure}

Figure~\ref{fig:sparse_bg} shows the slightly absurd spectral background estimate that results from employing a ``blank-sky'' spectrum extracted from a rather small 12.5 sq.\ arcsec region, as one might use to measure the core of a distant, compact galaxy cluster.
The raw spectrum contains just a handful of counts in the energy range of interest.
Binning the background spectrum itself to 1 count per channel produces an estimate with greatly degraded spectral resolution, while binning according to the same requirement applied to the source spectrum yields enormous statistical uncertainty in the background estimate at all energies.\footnote{One might additionally worry that a simple adaptive binning procedure like that used here, which ensures that all counts in a bin originated at the highest-energy channel it contains (see Figure~\ref{fig:sparse_bg}), could be problematic.}
This situation arises naturally if we wish to extract the most spatial and spectral information we can from a data set while insisting on employing an empirical background estimate from the same detector region (even from a much longer observation).

In principle, the ideal solution to the various problems described above is to use measurements of the backgrounds to develop a generative model (also called a forward model) of each of them, rather than relying on directly empirical estimates, whatever their source.
Such an approach allows us to explicitly leverage what has been learned about the astrophysical and instrumental backgrounds over decades of study.
In short, we actually have much more information about the background likely to be present in the region showed in Figure~\ref{fig:sparse_bg} than is provided by the tiny number of counts in the corresponding region of the blank-sky data set.
Both the SFG and CXB are subjects of astrophysical research in their own rights, and physically motivated models have therefore naturally been developed.
Knowledge of the QPB affecting a given instrument is necessarily empirical, but there have been significant efforts to understand its physical origin (e.g.\ \citealt{Marelli1705.04171, Marelli2012.02071, Ghizzardi1705.04173, Gastaldello1705.04174, Gastaldello2202.05286, Salvetti1705.04172, Fioretti2501.09724, Mineo2501.18346, Sarkar2405.06602, Schellenberger2503.04682}), and models consisting of continua and discrete emission features have been developed (e.g.\ \citealt{Bartalucci1404.3587, Suzuki2108.11234, Rossetti2402.18653}).

In short, with the community's accumulated 25 years of experience with the instrumental backgrounds in Chandra and XMM, and concurrent advances in our understanding of the astrophysical backgrounds, the time seems right to ask whether a complete generative model of these unwanted signals can now be employed, and what affect this might have on measurements of the ICM compared with other approaches.
This paper describes methods we are employing in concurrent work for defining and constraining such background models and applying them to particular Chandra or XMM observations.
These are not necessarily novel individually, and similarly motivated modeling of various background components can be found throughout the literature (e.g., \citealt{Snowden0710.2241, Eckert1408.1394, Sanders2106.14534}).
Indeed, our purpose here is not primarily to contribute substantively to the physical modeling of any of these backgrounds, but to synthesize the best practices for modeling them that are currently practical, in order to provide a useful benchmark.
In particular, few have taken advantage of the QPB models for ACIS produced by \citet{Bartalucci1404.3587} and \citet{Suzuki2108.11234}, and the practice of simultaneously fitting complete background and source models, as opposed to estimating and subsequently fixing background models, remains rare.
We demonstrate our approach with analyses of several galaxy clusters, spanning a range of redshift, mass and data quality, and
compare the results with those obtained using direct, empirical estimation of the backgrounds from blank-sky observations or source-free regions.
More prosaically, this paper serves as a reference for the methodology we are applying in a number of ongoing X-ray cluster studies.

Section \ref{sec:data} reviews the standard data reduction procedures we employ for Chandra and XMM.
In Section~\ref{sec:modeling}, we describe the models of the various background components, while Section~\ref{sec:application} covers the procedure for applying them to a particular target and set of observations.
Section~\ref{sec:examples} presents multiple example analyses with somewhat different goals, to demonstrate the impact of these methods.
We conclude in Section~\ref{sec:conclusion}.

\section{Data reduction} \label{sec:data}

This section summarizes the data reduction process for both Chandra and XMM.
Specific observations analyzed for this work are introduced on a per-cluster basis in Section~\ref{sec:examples}.
The Chandra data are compiled in Chandra Data Collection 491,\footnote{\dataset[doi:10.25574/cdc.491]{https://doi.org/10.25574/cdc.491}} while XMM data can be downloaded from the XMM-Newton Science Archive.\footnote{\url{https://www.cosmos.esa.int/web/xmm-newton/xsa}}

\subsection{Chandra}

Our procedure for reducing the Chandra data follows the recommendations in the Chandra Analysis Guide\footnote{\url{https://cxc.harvard.edu/ciao/guides/acis_data.html}} and was previously described by \citet{Mantz1502.06020}.
Apart from using newer versions of the Chandra analysis software ({\sc ciao} 4.16) and calibration files ({\sc caldb} 4.11.0), these steps remain unchanged.
In brief, level-2 event files are regenerated from the level-1 data using the {\sc ciao} script {\sc acis\_process\_events}, including standard grading and the use of VFAINT mode information where available.
We then use the {\sc ciao} tool {\sc wavdetect} to perform an initial identification of point-like sources, which may be time-variable AGN, and mask them while extracting light curves used to filter periods of enhanced background.
This filtering is generally accomplished using the {\sc lc\_clean} tool with default settings, except in rare cases where human intervention is required (typically when an extended background flare biases the mean rate even after sigma clipping).
The tool is run on data from a CCD that lacks strong cluster emission, since the mean absolute rate is assumed to reflect the background, generally chip I0 or S1.
When multiple observations of the same target exist, we also use the preliminary point source detections above to correct their relative astrometry.
This is of minimal importance for measurements of the intracluster medium, given Chandra's overall good pointing accuracy, but ensures that our final detection of point-like sources (Section~\ref{sec:cxb}) is as accurate as possible.

We also produce reprojected event files from the Chandra ``blank sky'' data sets and determine the ratio of the hard-band (9.5--12\,keV) rate in these data to that in a given science observation.
Our forward modeling procedure does not use such rescaled blank sky data, but they are necessary to compare our results with earlier methods.

After generating spectral files from Chandra data, we apply a correction to the Ancillary Response Files (ARFs) that encode the effective area as a function of energy.
The need for and nature of this correction is detailed in Appendix~\ref{sec:cal_correct}.
In brief, we find that a time-dependent modification is needed in order to obtain consistent results from observations of galaxy cluster A1795, in particular for data taken since $\sim2015$.
This correction is entirely empirical, and may not reflect any inaccuracy in the effective area calculations per se, but can be simply implemented as a modification of the ARF.
With some caveats discussed in the Appendix, the correction largely removes trends in the recovered gas temperature, metallicity and brightness that would otherwise be observed.

\subsection{XMM-Newton}

We reduce XMM data using the XMM-{\it Newton} Extended Source Analysis Software ({\sc xmm-esas}; {\sc sas} version 18.0.0), following the recommendations of \citet{Snowden0710.2241} and the {\sc xmm-esas} Cookbook.
After standard calibration and filtering (including automatic lightcurve filtering) of the raw event files and removal of MOS CCDs in anomalous states, lightcurves for each of the EPIC detectors were visually inspected for residual periods of enhanced X-ray background, and any such flares were manually removed.
For both images and spectra, the {\sc (e)sas} tools provide expectations for the QPB and out-of-time (OOT) event signals that we make use of in later sections; the QPB predictions are empirically based on data recorded with the filter wheel in the closed position, while the OOT predictions are obtained by resampling the detector positions of events from the observation in question along the readout direction.

We opted to skip the point source detection and relative astrometry correction steps in our XMM analysis, given that for all of the examples in this work we have complementary Chandra data that (due to its smaller PSF) is superior for finding point sources.
In principle, however, both steps could be done analogously to our Chandra procedure.

\section{Modeling of foregrounds, backgrounds and the ICM} \label{sec:modeling}

This section covers the models we use to describe each background component, as well as the ICM.
A few commonalities can be mentioned at the outset.
We use {\sc xspec} to evaluate each of the spectral models, as well as to compare them to the data.
Within {\sc xspec}, thermal emission is evaluated using the {\sc apec} model, with metal abundances scaled relative to the Solar table of \citet{Asplund0909.0948}.
Photoelectric absorption is evaluated using the {\sc phabs} model, with the equivalent absorbing hydrogen column density in a given direction provided by the HI4PI survey \citep{HI4PI1610.06175}.%
\footnote{Though the issue does not arise in this work, the neutral hydrogen-based estimates of the column density are known to be inaccurate for particularly dense lines of sight ($N_H \gtrsim 10^{21}$\,cm$^{-2}$; e.g.\ \citealt{Willingale1303.0843}); we recommend allowing the column density parameter to freely fit in such cases.}
Additional spectral models with more specific purposes will be mentioned below.
For both imaging and spectroscopy, we account for the XMM PSF using the symmetric Gaussian+King model developed by \citet{Read1108.4835}.
Whenever possible, we conservatively exclude the edges of the ACIS CCDs, whose exposure is affected by the Chandra dithering pattern, with masks extending $25''$ inward from the nominal chip edges in sky coordinates.
Finally, as described in Section~\ref{sec:cxb}, we limit Chandra analysis to chips I0--3 for ACIS-I observations and S1--3 for ACIS-S observations.

\subsection{Galactic X-ray foreground} \label{sec:sfg}

The soft X-ray foreground originating in the local Universe is typically described as the sum of emission from the local hot bubble and Galactic halo, potentially including contributions from unresolved stellar and extragalactic sources.
Modeling based on ROSAT data found that the typical foreground spectrum could be described as the sum of 2 thermal emitters with temperatures $\sim100$\,eV (one absorbed by Galactic hydrogen and one unabsorbed, reflecting a distribution in the distance to the emitters) and a hotter, absorbed thermal component with a typical temperature of $\sim200$\,eV (e.g.\ \citealt{Snowden1998ApJ...493..715S, Kuntz2000ApJ...543..195K}).
With the higher spectral resolution of XMM and Suzaku, additional contributions from SWCX (e.g.\ \citealt{Snowden0404354, Koutroumpa0805.3212, Ueda2209.01698}) and yet hotter thermal emission at $\sim700$\,eV (e.g.\ \citealt{Kuntz2008ApJ...674..209K, Das1909.06688, Das2106.13243, Huang2305.14484}) have been discussed.
The applicability of this set of model components has been broadly confirmed by eROSITA All-Sky Survey measurements \citep{Ponti2210.03133, Yeung2410.23345}.

Since a detailed study of these foreground emission is not the goal of this work, our strategy is to use a model with enough flexibility to describe the data, without necessarily including all the physical components that are present in detail.
Since Chandra and XMM differ radically in their sensitivity to $\lesssim1$\,keV photons, we take a different approach in each case.
As discussed below, for Chandra we use a simple model fit to ROSAT survey data from nearby on the sky, allowing more flexibility only if the Chandra data clearly require it.
The high sensitivity of the pn camera means that, in practice, we are always in the latter situation with XMM, and so we explicitly include a greater number of model components.

\subsubsection{Chandra}

\begin{figure}
    \centering
    \includegraphics[scale=0.175]{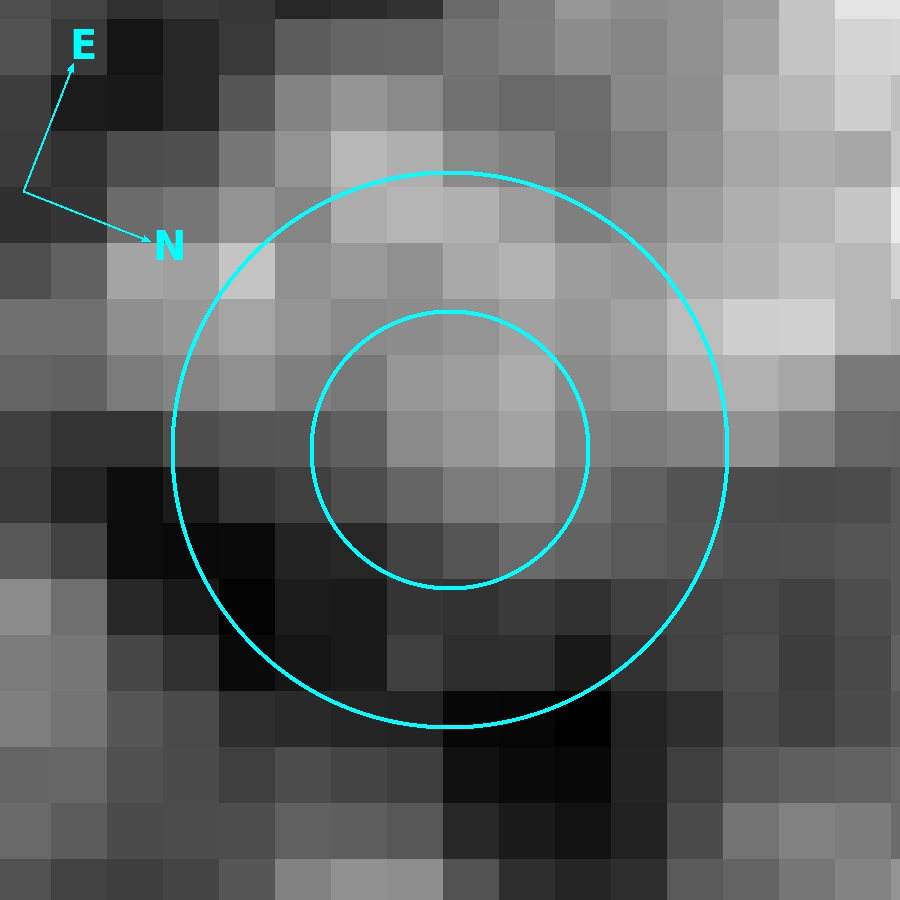}\\
    \includegraphics[scale=0.8]{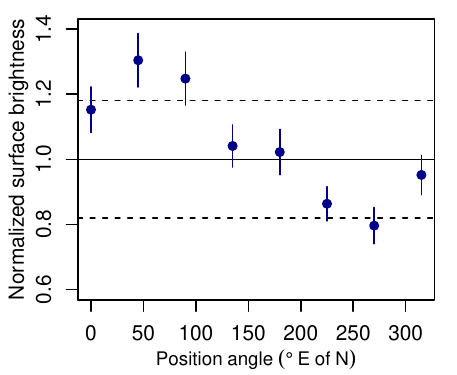}
    \caption{%
      Top:
      RASS diffuse background map in the direction of MACS\,J1423.
      Circles show the annulus spanning radii of $0.5^\circ$--$1^\circ$ from the cluster position on which we base the ROSAT SFG model.
      Bottom:
      SFG brightness within this annulus, relative to the mean, as a function of azimuth.
      Dashed lines indicate the standard deviation of these points, which are not consistent with a constant value.
    }
    \label{fig:rass}
\end{figure}

Since Chandra currently has relatively little sensitivity at soft energies,\footnote{Based on the data for A1795 that we analyze in Appendix~\ref{sec:cal_correct}, contamination on the optical blocking filter (e.g.\ \citealt{Plucinsky0209161, Plucinsky1809.02225}) has reduced the sensitivity of ACIS-S and ACIS-I since launch by factors of $\sim100$ and 10, respectively.} we obtain an independent constraint  from ROSAT All-Sky Survey (RASS) data from nearby regions of the sky.
Specifically, we used the {\sc heasoft} tool {\sc sxrbg} to obtain average X-ray background count rates across the six bands of the RASS diffuse background maps \citep{Snowden1997ApJ...485..125S, sxrbg1904.001} in an annulus between $0.5^\circ$ and $1.0^\circ$ from the nominal cluster position (these radii would need to be increased for particularly extended/low-redshift clusters; Figure~\ref{fig:rass}).
Following \citet{sxrbg1904.001},\footnote{\url{https://heasarc.gsfc.nasa.gov/Tools/xraybg}} we model the Galactic soft foreground as the sum of an unabsorbed thermal plasma component and two absorbed thermal components, where the absorbing equivalent \ion{H}{1} column density is that in the direction of the cluster.
The unabsorbed and one absorbed component are assumed to have temperatures of 0.1\,keV while the temperature of the third component is free, and all 3 are assumed to have solar metallicity.
A power-law component with a photon index of 1.45 is also included in the model, to account for the unresolved (at ROSAT resolution) CXB.
The free parameters of the model thus consist of the temperature of one unabsorbed thermal emitter, and normalizations for all 3 thermal components and the power-law model.
Apart from the power-law component, which is handled differently in later analysis, the best fit to the ROSAT data defines our fiducial model of the soft foreground.

As the foreground can vary on scales of $<1^\circ$, we perform a consistency test, repeating the above analysis for 8 azimuthally distinct subsets of the annular region.
Due to the smaller statistical power in these sub-annular spectra, the temperature parameter, power-law normalization and relative normalizations of the 3 thermal components are fixed in these fits, with only the overall thermal normalization free.
We then use a simple $\chi^2$ goodness of fit test to check whether the constraints on this normalization are statistically consistent with a constant value across all 8 sectors; if not (specifically, for a $p$-value $<0.05$), the scatter of the 8 normalizations is added in quadrature to the statistical uncertainty on the thermal normalization from the full annulus to determine the overall uncertainty on the foreground normalization from ROSAT.

When analyzing Chandra data, we similarly fix the temperatures and relative normalization of the thermal components, letting only their overall normalization vary.
(In practice, the 0.1\,keV components contribute so little in the energy band used in this analysis that they could be safely ignored.)
Typically, we fix all of the foreground temperature parameters at this stage, and marginalize over the normalization using a Gaussian prior defined by the ROSAT uncertainty, above.
In rare cases, the local foreground is bright enough that Chandra can provide comparable or better information on the normalization and/or temperature of the warm component.
In Section~\ref{sec:offcluster}, we will assess this and revise the foreground priors if warranted.

\subsubsection{XMM}

The larger sensitivity of XMM to soft X-rays makes it possible, and generally necessary, to more carefully model the SFG.
We retain the unabsorbed thermal emission model with temperature 0.1\,keV, now with its normalization fixed from the ROSAT analysis (it would be highly degenerate with the charge-exchange model introduced below), and the warmer absorbed thermal component (free temperature of, typically, $\sim0.2$\,keV).

In place of the 0.1\,keV absorbed thermal component, we include an AtomDB Charge Exchange ({\sc acx}) component \citep{Smith2012AN....333..301S, Smith1406.2037}.
We fit this model with a free equilibrium temperature and normalization, fixing the redshift to zero and the metallicity to solar, and using the default neutral helium fraction and the ``swcx'' flag (appropriate for emission within the heliosphere).
The quantum numbers of the charge-exchanged ions are approximated as a simple weighted distribution over the most populated shells and a Landau-Zener distribution over orbital angular momentum (model 7 in {\sc acx}).
Making this substitution allows for the possibility of different levels of SWCX contamination in different observations, and we find it to be statistically preferred in at least some cases.

We also allow for the possibility of an additional thermal component with temperature $\sim0.75$\,keV.
This component is not included by default, but its presence is tested for when spectrally analyzing regions free of cluster emission (Section~\ref{sec:offcluster}).
Given such a region, as always exists in this work, the foreground model is fit to the data without external priors.

\subsection{Point-like sources and the CXB} \label{sec:cxb}

Our approach to contamination by point-like sources is twofold: mask those that are detected and model those that are not.

\begin{figure}
  \centering
  \includegraphics[scale=0.32]{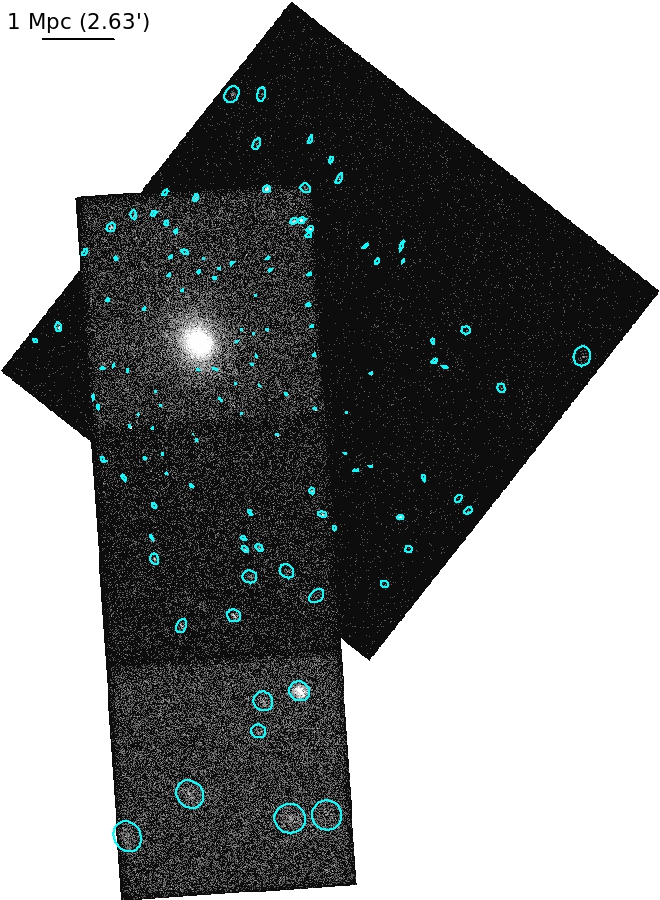}\\
  \includegraphics[scale=0.32]{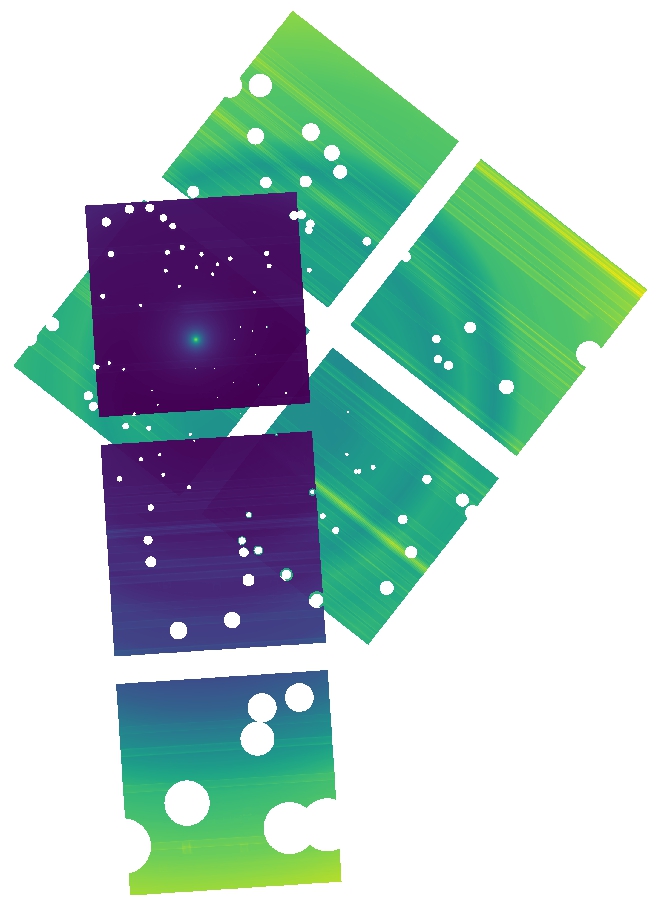}
  \caption{%
    Top:
    Chandra image of MACS\,J1423, combining 2 observations.
    Cyan ellipses show the sources identified by {\sc wavdetect}, after removing a spurious source at the cluster center. 
   These are not the masks we use in later analysis, which are ObsID-dependent.
    Bottom:
    Map of the expected CXB contamination due to undetected sources in the unmasked regions, from blue to yellow with increasing brightness.
  }
  \label{fig:cxb}
\end{figure}

For source detection, we have used Chandra data exclusively, although analogous steps could be implemented for XMM (with significantly reduced sensitivity, due to the larger PSF).
Having corrected the relative astrometry of all the Chandra observations of a target, we generate a combined image at native resolution in the 0.6--7.0\,keV band, and a corresponding exposure map at 2.25\,keV.
A final catalog of source detections is produced with {\sc wavdetect}, using scales of $2^{0,1,\ldots,5}$ pixels, and spurious detections associated with compact cluster cores or sharp features in the ICM are manually removed (Figures~\ref{fig:cxb}, top).
For cluster studies, we have been relatively aggressive in removing putative sources that are not readily apparent by eye within the region where the ICM emission is visible, though we note that one should be more cautious about such subjective determinations if the goal is to study the AGN population.
In all subsequent analysis of the Chandra data, each remaining source is masked within a circle centered on the {\sc wavdetect} position, with a radius corresponding to an enclosed energy fraction of 0.9997.
The mask associated with a given source may therefore have different radii in different observations of the same field, depending on their pointings and observation modes.

\begin{figure}
  \centering
  \includegraphics[scale=0.25]{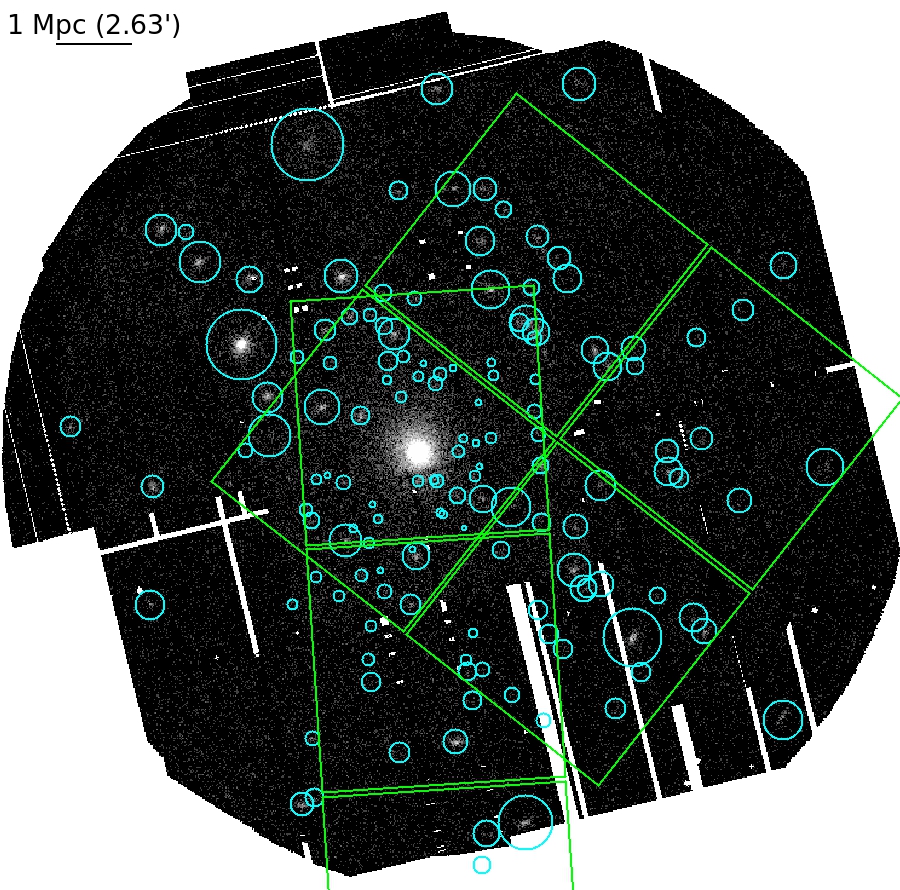}
  \caption{%
    MOS+pn image of MACS\,J1423 from a single XMM observation.
    Cyan circles show masks used to remove emission from point-like sources.
    Within the Chandra field of view (green squares), these are produced based on the {\sc wavdetect} source parameters and the XMM PSF model.
  }
  \label{fig:xmm_ptsrc}
\end{figure}

The Chandra-detected point source masks are resized for XMM analysis following the strategy of \citet{Mantz2006.02009}.
The {\sc wavdetect} count rate for each source was converted to an approximate 2--10\,keV flux assuming a power-law spectrum with a photon index of 1.4, and the radius of the corresponding mask chosen such that the maximum unmasked surface brightness according to the PSF model is $5\times10^{-19}$\,erg\,s$^{-1}$\,cm$^{-2}$\,arcsec$^{-2}$ in the 0.4--4.0\,keV band (with a maximum radius of $60''$; Figure~\ref{fig:xmm_ptsrc}).
This seemingly arbitrary value was originally chosen to produce masks that conservatively cover the visible emission.
The residual flux beyond the masks can nevertheless be computed given the Chandra flux estimates and XMM PSF model, and is typically negligible compared with other backgrounds.
In rare cases, we adopt a smaller mask for sources projected onto a cluster in order to avoid masking too much emission from the ICM (Section~\ref{sec:spt0459}; \citealt{Mantz2006.02009, Flores2108.12051}); here it naturally is more important to model the residual, unmasked emission.
In addition to the Chandra-detected sources, we visually identify and mask (to a comparable surface brightness) sources in the XMM data that were too faint to be detected at the time of the Chandra observations or fell outside of Chandra's field of view.
Note that, in this work, the regions identified for analysis of the ICM are entirely covered by Chandra.

\begin{figure}
  \centering
  \includegraphics[scale=0.9]{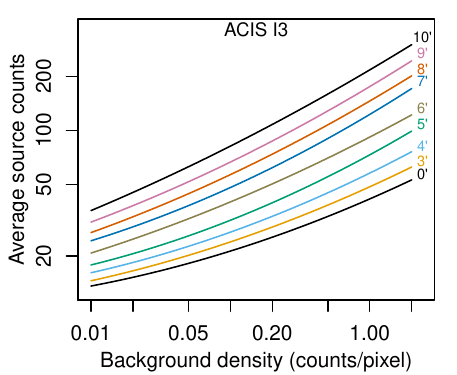}
  \caption{%
    Total source counts required for detection with 99 percent probability with {\sc wavdetect} as a function of the local density of background counts and the angular distance between the source and the aimpoint, based on Monte Carlo simulations with {\sc marx}.
    Source and background counts should both be interpreted as Poisson expectation values.
    These curves were determined for chip ACIS I3 and are applied to both I2 and I3; similar results are derived for use on chips I0--1 and S1--3.
  }
  \label{fig:ptsens}
\end{figure}

To predict the residual CXB signal from sources not detected by Chandra, we must first quantify the sensitivity limit of {\sc wavdetect} in a given Chandra observation.
To accomplish this, we employed the ray-tracing code {\sc marx} to simulate point-like sources at different positions in the ACIS-I and ACIS-S arrays.
Note that {\sc marx} does not have the ability to simulate events on the ACIS-I chips when the aimpoint is on ACIS-S or vice versa; the CCDs for which we consequently cannot produce a model for the unresolved CXB are removed from all later analysis.\footnote{This is not necessarily a great loss, given the significant vingetting and large PSF on chip S2 in ACIS-I configuration, and the fact that chips I2 and I3 are tilted away from the focal plane, and thus out of focus, in ACIS-S configuration.}
We empirically quantify the probability of {\sc wavdetect} to detect a given source as a function of the source's angular distance from the aimpoint; its flux, parametrized by the Poisson mean of the number of counts it contributes to a simulated image; and the density of background counts, parametrized by a uniform Poisson mean per pixel.
Given that the variation of PSF size with position depends on the tilt of the CCDs \citep{Garmire2003SPIE.4851...28G}, we perform separate suites of simulations for chips I0--1, I2--3 and S1--3.
For each of these subsets of the detector, the outcome of our simulations is a set of relatively smooth, interpolatable curves describing the source flux corresponding to 99 percent completeness from {\sc wavdetect} as a function of off-axis position and background surface brightness (all parametrized as above).
Figure~\ref{fig:ptsens} shows this family of sensitivity curves as determined from aimpoints simulated on chip I3; parameters defining such curves for all CCDs are given in Appendix~\ref{sec:ptsens}.
In each case, we find that the sensitivity curve varies negligibly between $0'$ and $2'$ off axis.

In the context of point source detection, the ``background'' surface brightness density consists of all of the foreground and background components discussed in this section (including all \emph{other} undetected point sources) \emph{and} the ICM or other target of observation.
An accurate prediction for the unresolved CXB can thus only be made following a preliminary imaging analysis of the observed field, as we perform in Section~\ref{sec:imaging}.
Once such a model of the ``background'' as a function of position is available, we can compute a corresponding map of the sensitivity limit for point source detections, converting the limiting mean number of counts per pixel to a limiting unabsorbed surface brightness in the 2--10\,keV band, assuming a power-law spectrum with photon index 1.7.
Following \citet{de-Vries2211.07680}, we then integrate the AGN luminosity function model of \citet{Miyaji1503.00056} up to this limit as a function of position to form a predicted map of emission from undetected CXB sources.
This map can be converted to absorbed surface brightness in an appropriate energy band for subsequent imaging analysis, or integrated in a region of interest to normalize a CXB model used in spectral analysis (Figure~\ref{fig:cxb}, bottom).
When modeling the unresolved CXB, primarily consisting of sources much fainter than the sensitivity limit, we assume a shallower photon index of 1.4.

\subsection{Quiescent particle-induced background}

\subsubsection{Chandra} \label{sec:chqpb}

Our approach to modeling the QPB in Chandra data is based on the {\sc mkacispback} code of \citet{Suzuki2108.11234}.
Those authors used data taken with ACIS when it was in a stowed position, unexposed to the sky, to develop a QPB model comprising continuum and emission line components for both FAINT and VFAINT-mode data, including the dependences on detector position and (for ACIS-S) variations in spectral hardness.

We implemented minor modifications the \citet{Suzuki2108.11234} code, as follows.\footnote{These changes have now been incorporated into the official repository at \url{https://github.com/hiromasasuzuki/mkacispback}.}
First, when combining the spectral models defined for different parts of the detector to produce a composite prediction for a given region, we use a flat weight map (i.e., area weighting) rather than a weight map defined by the hard band (9.0--11.5\,keV) counts in the observation of interest.
This is well motivated, and better behaved (does not introduce Poisson fluctuations) when considering small regions that contain relatively few counts.
Second, when normalizing the QPB model to account for gross temporal variability, rather than fitting the normalization to the observed spectrum in the hard band, we simply renormalize based on the total count rate in that band.
This change appeared to improve the comparison to observed data over a wider energy range, albeit at a relatively minor level.
Finally, we introduced flexibility to customize the definition of the ``hard band'' used for a given data set in specific cases, to accomodate observations where a more strict telemetry limit was employed.

In practice, we first generate a whole-chip QPB model for each active CCD, maximizing the signal available, and minimizing the Poisson noise, for the hard-band normalization step.
When producing QPB models for smaller spectral regions located on a given CCD, we directly apply the normalization obtained from the whole chip, rather than one derived from the smaller region.
In imaging and spectral analysis, we marginalize over the Poisson uncertainty in each CCDs normalization, as well as a correlated, per-observation Gaussian systematic allowance at the 5 percent level.

\subsubsection{XMM}

For XMM analysis, filter-closed data perform the same function as the Chandra stowed data sets.
Furthermore, {\sc sas} has long provided ``background'' data sets tailored to a given observation by resampling filter-closed data taken under comparable conditions, as determined from events recorded in the unexposed ``corner'' regions of the EPIC detectors, excluding energy ranges associated with certain fluorescence lines.
Standard {\sc sas}/{\sc esas} processing produces such a tailored background file for each science image and spectrum.
These background files do not constitute a generative model in the sense that we advocate for in Section~\ref{sec:intro}, in contrast to the ACIS QPB model of \citet{Suzuki2108.11234}, which synthesizes the available background-only data to provide a physically motivated prediction for the background in a given observation.
Nevertheless, with the few compromises described below, we can treat them as such in practice.

The SAS background files retain Poisson fluctuations from the original filter-closed data that do not represent our expectation for the mean background in a given pixel or channel.
We find that a small amount of Gaussian smoothing, at or below the PSF scale or spectral resolution, is sufficient to regularize the background such that Poisson fluctuations in the original data do not unduly influence the results.
In the imaging context, edge effects are typically negligible, provided that unobserved pixels are excluded from the averaging rather than being included with values of zero.
For pn spectra, smoothing with a 25\,eV (standard deviation) kernel does cause the smoothed spectrum to systematically depart from the original, but only at energies below the minimum of 0.4\,keV that we use for analysis.

An exception occurs for spectroscopy of sufficiently small regions, where the resampled events are simply too few for the shape of the background spectrum to be meaningfully reproduced, even after smoothing.
For our purposes, this situation arises for annular regions used to constrain ICM profiles, either for a subset of annuli near the cluster center or, for sufficiently distant/compact clusters, for all annuli.
Fortunately, assuming that the XMM observations intentionally targeted the cluster, these regions will be near the aimpoint, where the QPB is relatively uniform.
Our solution to this case is to create an additional spectrum for a circle of radius $2'$ about the aimpoint/cluster center, whose background model (appropriately scaled) is applied to all cluster annuli with smaller radii.
Annuli with radii larger than this are typically wide enough that their background spectra are well sampled, though if that were not the case then a similar strategy could be employed to provide a local but well sampled background model.

For image analysis, we apply similar priors to the Chandra case, marginalizing over a 5 percent systematic allowance for the normalization of the QPB in a given observation, with the relative normalizations of MOS1, MOS2 and pn independent for each observation.
For spectroscopy, given that the EPIC corner data allow the empirical QPB model to be tailored to a given observation, in principle accounting for gross variations in spectral shape, we do not adopt further informative priors.
Instead, we include high energy data, up to 12\,keV for MOS and 15\,keV for pn, that directly constrain the QPB normalizations.

\subsection{Out-of time events}

Events that arrive during readout (or while charge is being transfered to a frame store) are assigned incorrect detector positions and consequently receive incorrect gain/charge transfer inefficiency corrections, functionally becoming an additional source of background.
This contamination is typically significant only for the XMM pn detector, due to its combination of high sensitivity and relatively slow readout for the Full Frame mode most commonly used for cluster observations; in this mode, OOT events comprise 6.3 percent of the observed signal.
Bright enough sources can still leave clear ``readout streaks'' in ACIS or MOS data, but these are relatively rare in cluster observations, and typically it is sufficient to mask any such visible streaks.
In this work, we explicitly model OOT events only for pn data.

As with the QPB, {\sc sas} provides OOT images and spectra estimated from resampling.
In this case, the position of the science observation's events themselves in the readout direction are resampled, and the resulting event file is processed identically to the original observation.
We can use these estimates as the basis for a model of the OOT signal, similar to our strategy for the QPB, with a few adjustments.

The first clear difference between the OOT and QPB backgrounds is in their spatial dependence.
Since the OOT events are misplaced along the readout direction but not the perpendicular direction, smoothing an OOT image with a symmetric kernel would be a mistake (unless the kernel width were quite small).
In principle, one could smooth only in the readout direction, or even average the OOT image in each column, although the latter would neglect any position-dependent miscalibration effects.
In practice, since the OOT signal is generally smaller than the  other backgrounds that contribute to it, we have found that using the original OOT images without smoothing (i.e., with unphysical Poisson fluctuations intact) works well enough.
However, we note that the use of inappropriate calibration changes the reconstructed energy of an OOT event compared with the original, so there may be circumstances and particular energy ranges where the OOT background is dominant, and where more care would be needed.

The other consideration is that the data set to be resampled to estimate the OOT signal is limited to the original science observation, and may therefore be significantly smaller than the filter-closed data used to estimate the QPB.
The upshot is the OOT spectra produced for small regions can be even less populated, making them a poor representation of the underlying signal.
We address this by fitting a continuous model to the OOT spectra rather than simply smoothing them.

Physically, the OOT signal consists of events from the ICM as well as all the backgrounds discussed above.
However, the miscalibration of the OOT events makes the resulting spectrum nontrivially different from a simple superposition of these components.
Broadly speaking, we expect to see a continuum attributable mostly to the SFG and QPB, as well as particularly bright fluorescence lines, though the line centers and widths may be shifted.
Beyond this basis, our aim is simply to produce an empirical description of the estimated OOT spectra.

\begin{figure}
  \centering
  \includegraphics[scale=1]{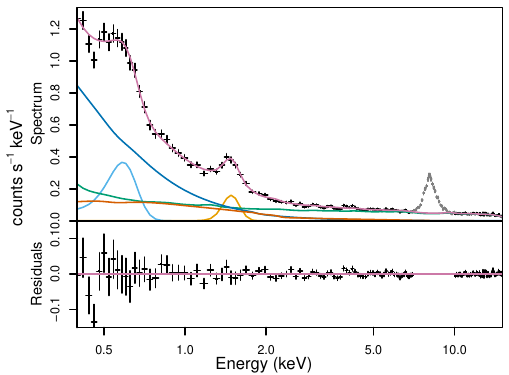}
  \caption{%
    OOT spectrum corresponding to the cluster-free region of 1 XMM observation of MACS\,J1423 (see Figure~\ref{fig:annuli}).
    Curves show the various model components described in the text, as well as their sum.
    Points shown in gray are not included in the fit.
  }
  \label{fig:oot}
\end{figure}

Correspondingly, we construct an OOT model out of the following components: 2 absorbed power laws, 2 Gaussian lines, and the region-appropriate QPB model (Figure~\ref{fig:oot}).
In this instance, we allow the power-law normalizations to take negative values in order to best fit the continuum.
One of the power laws and Gaussians is folded through the instrument response, while the other components are not.
We fit over the 0.4--15.0\,keV energy range, excluding 7--10\,keV, since our science analysis avoids the range dominated by Cu fluorescence in any case.
For regions smaller than 16 square arcmin, it is generally not possible to fit all parameters of the model; in this case we fix the folded Gaussian to have a center of 0.61\,keV and standard deviation of 48\,eV, and the unfolded Gaussian to have a center of 1.55\,keV and standard deviation of 72\,eV, values that appear to work well in general.
After fitting, this complex OOT model is converted to a table model with a fixed spectral shape, normalized appropriately by the ratio of the readout time to the frame time.
Given the small amplitude of the OOT signal compared with other backgrounds, we have chosen to fix its normalization rather than marginalize over it.

\subsection{Spectral modeling and deprojection of the ICM}

Our strategy for spectrally modeling the cluster follows that of \citet{Mantz1402.6212, Mantz1509.01322}.
Given spectra extracted in a series of circular annuli covering the cluster (see Section~\ref{sec:centering}), the ICM is modeled as a series of concentric, isothermal, spherical shells (each represented by an absorbed thermal emitter), with inner and outer radii corresponding to those of the annuli.
In practice, this procedure allows the densities in each shell to be constrained individually, while temperatures and metallicities must be linked between groups of adjacent shells.

For Chandra spectra of cluster regions, to limit the impact of background modeling uncertainties, we  constrain the range of energies in our analysis to those where the expected source spectrum, at a given radius, exceeds the background (0.6--7.0\,keV at widest).
This step requires an initial estimate of the cluster temperature profile; however, the steeply falling effective area at energies $\gtrsim2$\,keV means that the allowed energy ranges are not particularly sensitive to these initial temperature estimates.
At large radii/low ICM surface brightness where the allowed range would be narrower than 1\,keV, we transition from using spectral information to using only the integrated 0.6--2.0\,keV surface brightness (or 1.0--2.0\,keV for recent data, for which the 0.6--1.0\,keV sensitivity has significantly degraded).
A similarly conservative approach could be used for XMM, though in the data looked at so far we have yet to see a case where the background model failed to provide a consistently acceptable description across all energies and spatial regions.

The geometric projection of the 3D model onto 2D annuli in the plane of the sky \citep{Kriss1983ApJ...272..439K} is handled by the {\sc projct} model in {\sc xspec} for Chandra data.
For XMM, the additional mixing due to the PSF cannot be neglected, and so we implement the model more explicitly as a linear combination of thermally emitting components.
Following \citet{Mantz2006.02009}, we employ the PSF model of \citet{Read1108.4835} and compute the PSF mixing between annuli based on a $\beta$-model fit to the surface brightness distribution of the ICM from the Chandra data (see Section~\ref{sec:imaging}).
In both cases, the projected emission from 3D radii larger than the extent of the final cluster annulus is accounted for post-facto by assuming a locally $\beta$-model profile (essentially equivalent to a power law at large radii).
This final step often has a tens-of-percent impact on the density of the largest-radius shell modeled, but negligible effect at smaller radii (see \citealt{Mantz1402.6212}).

\section{Analysis of a given field} \label{sec:application}

The first step in analyzing the data for a given cluster and the surrounding field is to obtain a baseline SFG model from offset ROSAT data, following Section~\ref{sec:sfg}.
We then proceed with imaging and spectral analysis of the Chandra and/or XMM data as described below.

\subsection{Determining the unresolved CXB component} \label{sec:imaging}

As described in Section~\ref{sec:cxb}, a preliminary image model is needed in order to compute the residual CXB normalization as a function of position, after masking point-like sources identified in Chandra data.
We use images in the 0.6--7.0\,keV band and 2.25\,keV exposure maps, matching those employed for source-finding, but binned by a factor of 4 for the sake of efficiency.
Multiple observations are fit in parallel, using the correct Poisson likelihood for the number of counts in each pixel and masking point sources detected in Section~\ref{sec:cxb}.
The image model at this stage consists of the CCD-specific QPB contribution, a spatially uniform, vignetted component (comprising the SFG and CXB) and a beta model representing the ICM.
We convert the SFG model from Section~\ref{sec:sfg} to an expectation for the foreground surface brightness, and adopt this value minus 3 times its assigned uncertainty as the minimum for the combined SFG+CXB model, adopting a log-uniform prior over larger values.
The best-fitting model is used as the ``background'' to determine the sensitivity limit for detecting CXB sources throughout the unmasked parts of the field (Section~\ref{sec:cxb}).

With an expectation for the CXB now in hand, we turn to testing the complete foreground and background model in the next section.

\subsection{Testing the foreground and background models} \label{sec:offcluster}

\begin{figure}
  \centering
  \includegraphics[scale=0.32]{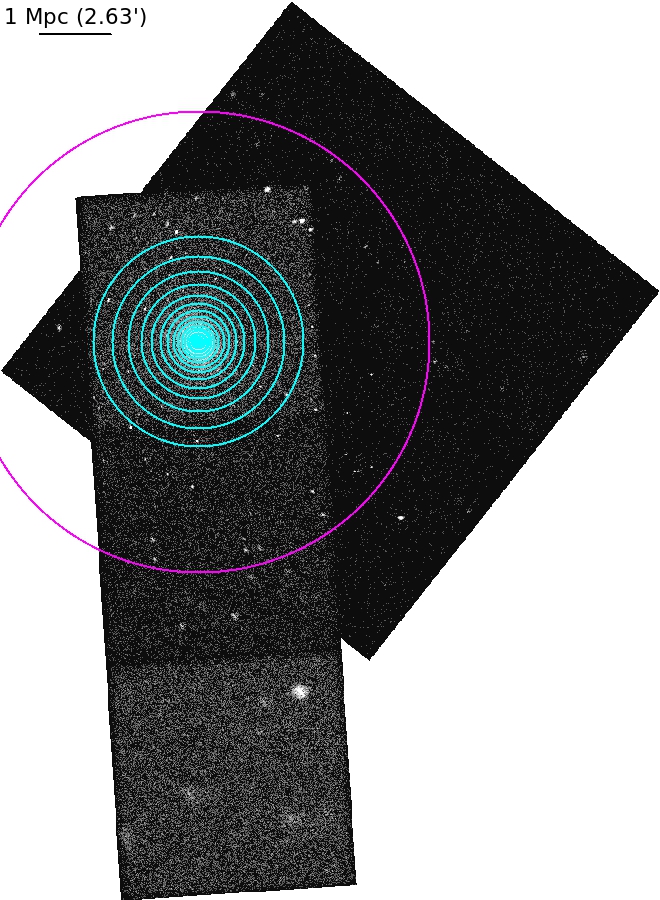}\vspace{5mm}\\
  \includegraphics[scale=0.25]{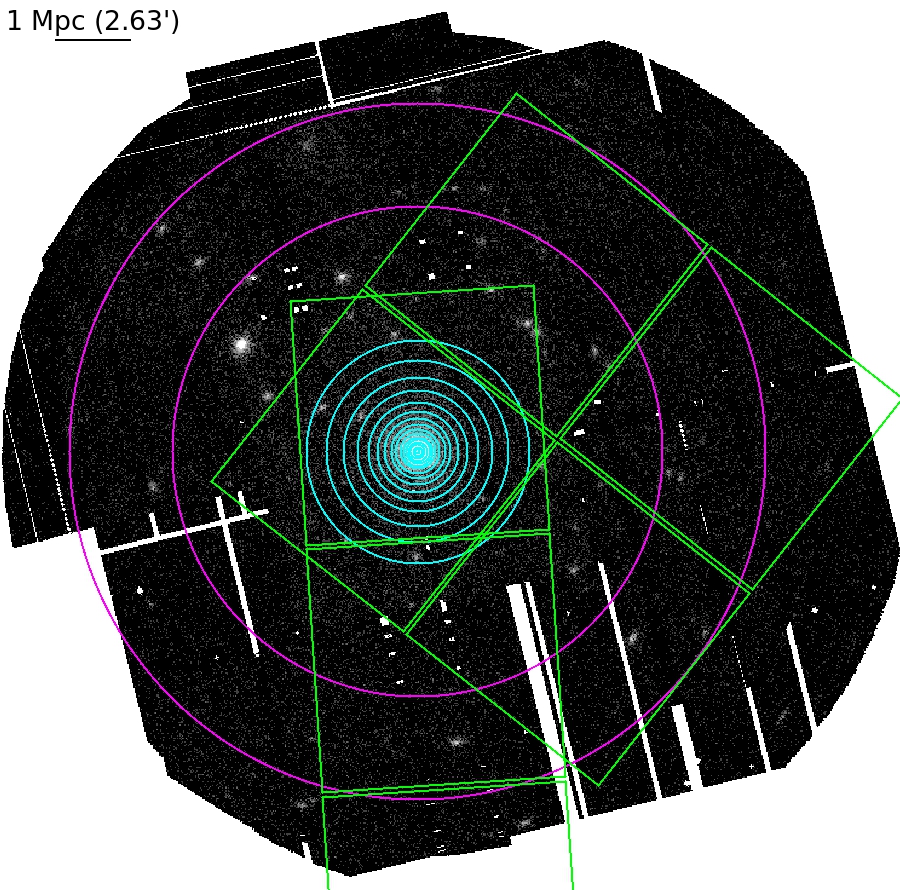}
  \caption{%
    Chandra (top) and XMM (bottom) images of MACS\,J1423.
    Annular regions used to for spectral analysis of the ICM are shown in cyan.
    The cluster-free region that is simultaneously used to constrain foreground and background model parameters consists of all available area outside the magenta circle for Chandra, and the interior of the magenta annulus for XMM.
    For reference, the Chandra field of view is shown as green squares over the XMM image.
  }
  \label{fig:annuli}
\end{figure}

\begin{figure*}[ht!]
    \centering
    \includegraphics[scale=1]{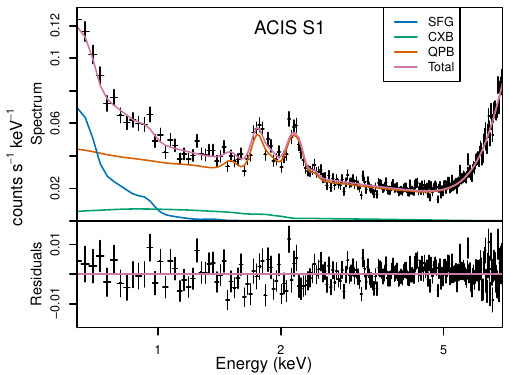}
    \hspace{5mm}
    \includegraphics[scale=1]{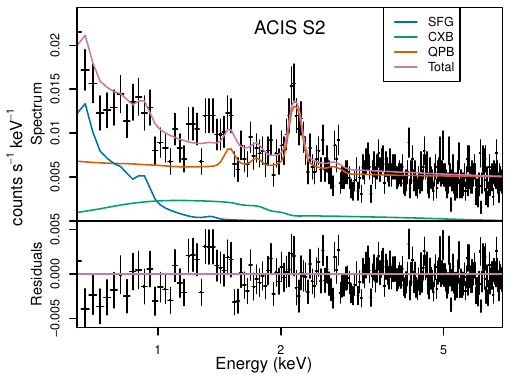}\vspace{5mm}\\
    \includegraphics[scale=1]{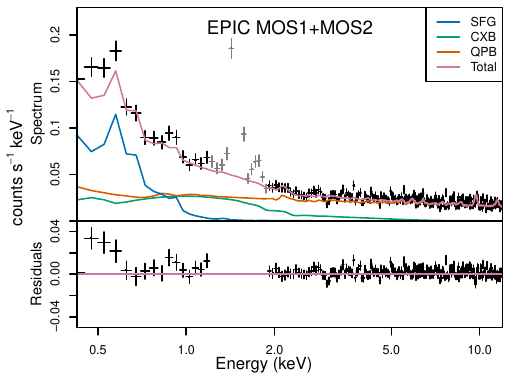}
    \hspace{5mm}
    \includegraphics[scale=1]{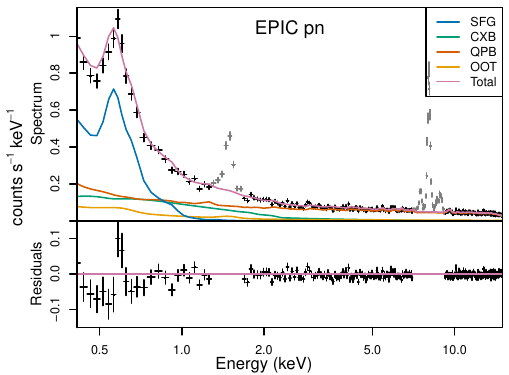}
    \caption{%
      A subset of the cluster-free spectra used in the analysis of MACS\,J1423, specifically from Chandra ObsID 4195 (top row) and XMM ObsID 0720700301 (bottom row).
      Curves show the various models and their sum.
      Gray points shown in the bottom panels are not included in the fits.
      The disagreement between MOS and pn at $\sim0.5$\,keV is a consistent feature in the XMM data used in this work.
    }
    \label{fig:offcluster}
\end{figure*}

When an observation contains regions free of cluster emission, they can be used to verify the foreground and background models, and to test whether additional complexity is needed.
For practical purposes, we define ``cluster-free'' to mean at a distance $>3\,r_{500}$ (approximately $2\,r_{200}$) from the cluster center, where the center can be approximate and $r_{500}$ can be taken from the literature/previous analysis or crudely estimated from a scaling relation, since it need not be particularly accurate.
For all the clusters presented in this work, we have estimates of both from previous work \citep{Mantz1606.03407, Mantz2111.09343}.
For XMM, we also limit the analysis to be within $12'$ of the aimpoint, as this provides ample data while avoiding the significant vignetting at larger radii (Figure~\ref{fig:annuli}).
Example spectra and fitted models are shown in Figure~\ref{fig:offcluster}.

For Chandra data, in addition to verifying that the adopted priors on the CXB and QPB components acceptably describe the data, we can check whether the soft foreground model obtained from ROSAT data 30--60$'$ away from the cluster is sufficient.
Within the Chandra bandpass of 0.6--7.0\,keV that we use for analysis, only the $\sim0.2$\,keV (or warmer) thermal components of the model can plausibly be detected, and then only if it is brighter than typical.
Therefore, we first compare the goodness of fit of a model where the overall normalization of the SFG is free (with a uniform, positive prior) with the baseline model where it is constrained by a Gaussian prior determined from the ROSAT data.
In both cases, the temperatures and relative normalizations of the thermal emitters are fixed to the ROSAT best fit.
If the former is a statistically significant improvement, defined here as a reduction in the Bayesian information criterion equivalent to a significance of 0.05, we proceed to test whether the fit is improved by allowing the temperature of the warm component to be free.
The most complex of these models justified by the data is adopted for subsequent analysis, where the cluster-free spectra will be fit in parallel with cluster spectra.

For XMM, we consider the energy range 0.4--15.0\,keV for pn (excluding fluorescence lines at 1.3--1.65 and 7.0--9.2\,keV) and 0.4--12.0\,keV for MOS (excluding 1.2--1.9\,keV).
Observations that contain residual soft proton contamination (i.e., neither removed by lightcurve filtering nor accounted for by the QPB prediction) can be identified by looking for residuals between the pn data and corresponding QPB spectra at the highest energies ($\sim9$--15\,keV); none of the data used in this work suffer from such contamination.
The cluster-free regions available in XMM typically extend well beyond the area for which we have a CXB model available from Chandra observations of the same cluster, so we allow the normalization of the CXB to fit freely in this case.
The baseline SFG model includes the cool/unabsorbed and warm/absorbed thermal components fit from ROSAT, and SWCX emission whose temperature and normalization are independent in each observation.
From here, we follow an analogous procedure to decide what parameters should be free for a given set of observations.
Invariably, we have found that freeing the temperature and normalization of the warmer emitter results in an improved fit; we also test whether including a 0.75\,keV thermal emitter, making the CXB surface brightness independent in different observations and/or freeing the CXB spectral index improve the fit.
In the case of the CXB model, these decisions apply only to the cluster-free region and not to CXB contamination of cluster spectra, since in this work we always have a Chandra-based model for the residual CXB for cluster regions.

\subsection{Definition of cluster centers and annular regions} \label{sec:centering}

Having checked the appropriateness of the foreground and background models, and potentially revised the prior on the SFG surface brightness, we continue analysis of the Chandra images, specifically binned by a factor of 2 and in the 0.6--2.0\,keV band, where the QPB is relatively small compared to the focused X-ray signal.
We first fit a model consisting of the SFG (uniform and vignetted), CXB (non-uniform and vignetted), QPB (non-uniform and not vignetted) and a beta model describing the ICM.
The non-ICM components of this fitted model are interpreted as background for the purpose of finding the center of the cluster using the Symmetry-Peakiness-Alignment (SPA) algorithm applied to unbinned images (\citealt{Mantz1502.06020}; in previous work, the Chandra blank sky data served as a background estimate instead).
For relaxed clusters, the result is sometimes adjusted at the arcsecond level to lie at the center of the cool core. 
Fixing this as the center of the cluster model(s), we next test whether including a second $\beta$ component to describe the ICM substantially improves the goodness of fit, and retain the double-$\beta$ model if so.
From this final model fit, we then define a set of annuli, approximately logarithmically spaced, spanning from the cluster center to the radius where the cluster signal represents a $2\sigma$ excess over the background in the 0.6--2.0\,keV band (Figure~\ref{fig:annuli}). 
Spectra extracted from these regions, along with cluster-free regions if available, are used in the analysis described in Section~\ref{sec:deprojection}.

An analogous process can be followed for XMM data to independently determine a set of annuli for deprojection.
Given the size of the PSF, we limit these annuli to be no less than $5''$ wide (inner to outer radius).

\subsection{Final spectral analysis} \label{sec:deprojection}

With the ingredients described in the preceding sections, we are now in a position to fit for the properties of the ICM as a function of radius while simultaneously fitting or marginalizing over the various foregrounds and backgrounds.
To recapitulate, the data will consist of spectra extracted from a series of concentric annuli, as well as, when available, a large region in which cluster emission is assumed to be absent.
The model accounts for the ICM (in the annular regions), Galactic soft emission, the unresolved CXB, particle-induced events and OOT events (for pn data).
Free parameters include those describing the ICM, the normalizations of the CXB and QPB backgrounds (on a per-observation basis) and parameters of the SFG.
The latter depend on what SFG components are justified by the data in a given field, but at a minimum the overall normalization is always marginalized over, as well as the temperature and normalization of the SWCX signal (per-observation, for XMM data).
The specific priors used for different foreground and background parameters are described in the previous sections; for those not explicitly mentioned, uniform priors are employed.

As in previous work, we use {\sc xspec} to evaluate the likelihood of the data given the model, passing this to the Markov Chain Monte Carlo code {\sc lmc} to estimate the posterior distribution.
In this case, the likelihood is determined from the classical Cash statistic \citep{Cash1979ApJ...228..939} without the need to bin the data in energy space.

\section{Examples and performance} \label{sec:examples}

In this section, we apply the modeling described above to the analysis of several clusters, and compare the results with those obtained using our previous methods.
As a guide, in this section we present analyses of
\begin{enumerate}
\item MACS\,J1423.8+2404 (hereafter MACS\,J1423), an intermediate-redshift, cool-core cluster, observed to similar depth with Chandra and XMM;
\item WARP\,J1415.1+3612 (hereafter WARP\,J1415), a relatively faint $z>1$ cluster with deep Chandra data, hosting an exceptionally high-metallicity core and a correspondingly steep metallicity gradient;
\item MACS\,J0717.5+3745 (hereafter MACS\,J0717), a complex merger where extremely high temperatures ($>20$\,keV) have been reported from Chandra data, for which we constrain a 2D temperature map rather than performing a deprojection; and
\item SPT-CL\,J0459$-$4947 (hereafter SPT\,J0459), a $z>1.7$ cluster with shallow Chandra and very deep XMM data.
\end{enumerate}
For the first and last of these, we analyze both the Chandra and XMM data, using the former to inform the latter in the limited ways described above.
For MACS\,J1423, we compare the results from the 2 observatories, while the Chandra data for SPT\,J0459 are too shallow to do so meaningfully.
In the remaining cases, we consider only the Chandra data.
These case studies span a range in cluster brightness, angular extent, temperature and morphology, and our results are suggestive of the benefits expected from careful modeling of all the measured signals in various circumstances.
We emphasize, however, that more extensive analysis of a larger cluster sample would be needed to draw firm conclusions about the general impact of this approach.

When we compare results using the methods described above with those where backgrounds are estimated directly from blank sky data, the same science spectra are used.
That is, both analyses fit the same spatial regions (including point source masks) and energy ranges, as determined in the previous sections, even though some relevant details have changed compared with our earlier work.
We also note that the impact of the background modeling choices is necessarily muted by our generally conservative approach, in particular avoiding drawing conclusions from data where the cluster signal is smaller than the expected background.

\subsection{Chandra+XMM: MACS J1423.8+2404}

Our first test case for the methods described in this work is MACS\,J1423, a redshift 0.539 galaxy cluster discovered in the ROSAT All-Sky Survey and first cataloged in the MAssive Cluster Sample (MACS; \citealt{Ebeling0703394}) that has been observed to similar (and useful) depth with both Chandra and XMM (Table~\ref{tab:macs1423obs}).
Figures~\ref{fig:rass}--\ref{fig:xmm_ptsrc} and \ref{fig:oot}--\ref{fig:offcluster} illustrate various intermediate stages of our analysis for this cluster;
in particular, Figure~\ref{fig:offcluster} includes a subset of the cluster-free spectra used to verify the adequacy of the foreground and background models.
Worth noting for this cluster is the relatively small angular extent of the X-ray emission and its sharp surface brightness peak, both of which increase the salience of the PSF modeling for XMM.
In addition, MACS\,J1423 lies behind a region of relatively bright and spatially variable Galactic emission (Figure~\ref{fig:rass}), such that the SFG is clearly visible in the Chandra data (at least in this relatively early ACIS-S data, which predates the significant reduction in soft response).
Following Section~\ref{sec:offcluster}, the overall normalization of the SFG model used in the Chandra analysis is not constrained by a ROSAT-based prior.

\begin{figure*}
  \centering
  \includegraphics[scale=1.0]{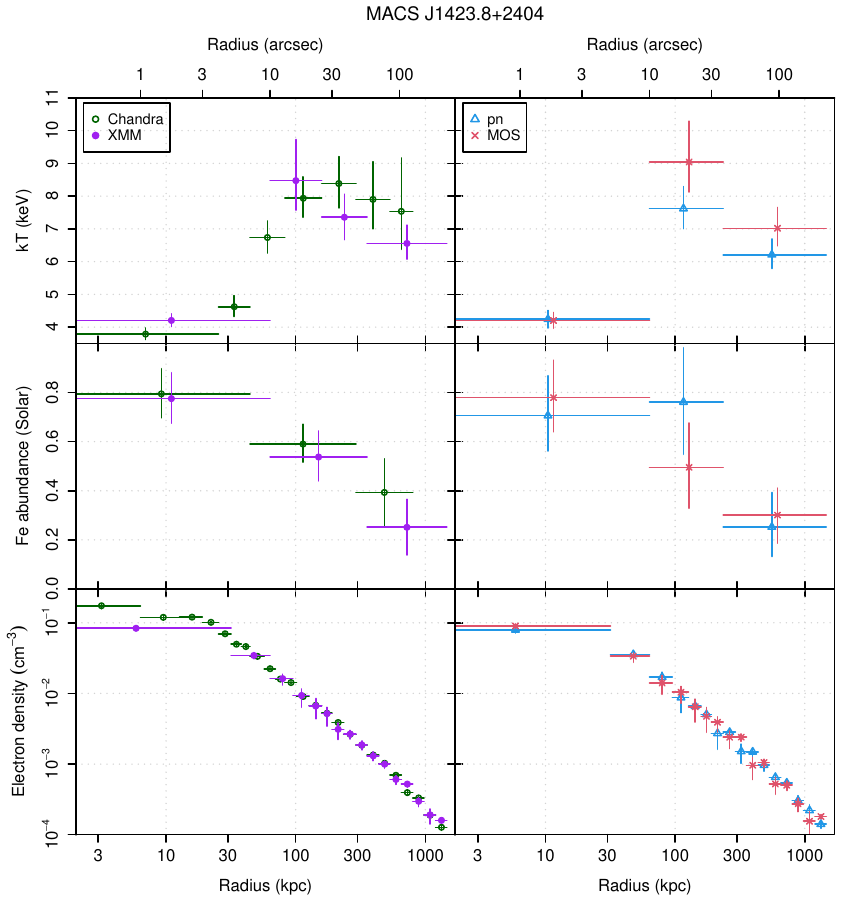}
  \caption{%
    Left: Deprojected temperature, metallicity and density profiles of MACS\,J1423 from our Chandra and XMM analyses.
    Right: Profiles using the pn or MOS detectors of XMM separately.
    We fit fewer temperatures for the latter comparison, due to the smaller statistical power of the MOS data.
  }
  \label{fig:m1423_chxmm}
\end{figure*}

\begin{table}
  \caption{Observations of MACS\,J1423}
  \hspace{-15mm}
  \begin{tabular}{crccD}
    \hline
    Obs. & \multicolumn{1}{c}{ObsID} & Det. & Date & \multicolumn{2}{c}{Clean exp.}\\
    &&&& \multicolumn{2}{c}{(ks)}\\
    \hline
    \decimals
    CXO & 1657 & ACIS-I & 2001-06-01 & 18.5 \\
    CXO & 4195 & ACIS-S & 2003-08-18 & \hspace{1ex}115.6 \\
    XMM & 0720700301 & MOS1 & 2014-01-29 & 32.0 \\
    \ldots & \ldots & MOS2 & \ldots & 35.1 \\
    \ldots & \ldots & pn & \ldots & 20.8 \\
    XMM & 0720700401 & MOS1 & 2014-01-31 & 22.8 \\
    \ldots & \ldots & MOS2 & \ldots & 22.6 \\
    \ldots & \ldots & pn & \ldots & 19.2 \\
    \hline
  \end{tabular}
  \label{tab:macs1423obs}
\end{table}

\begin{figure*}
  \centering
  \includegraphics[scale=1.0]{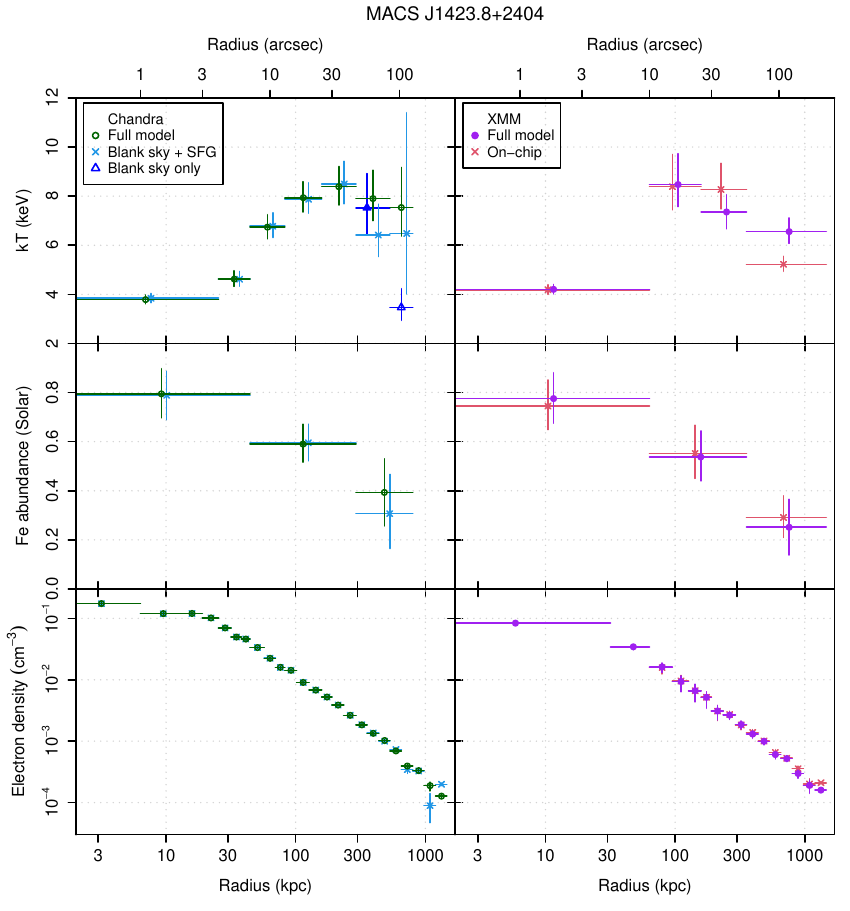}
  \caption{%
    Comparison of MACS\,J1423 profiles obtained from Chandra (left) and XMM (right) using the methods described in this work with blank-sky based approaches.
  }
  \label{fig:m1423_oldbg}
\end{figure*}

 Figure~\ref{fig:m1423_chxmm} (left column) shows the ICM temperature, metallicity and density profiles determined independently from Chandra and XMM.
We are surprised and delighted by the level of agreement on display.
The XMM temperatures represent a compromise between the MOS and pn detectors, which have been reported to be systematically offset for the warmer temperatures present outside the cluster core (e.g.\ \citealt{Schellenberger1404.7130}\footnote{This work used a blank-sky background description for both Chandra and XMM, rather than the forward modeling described here. While their restriction to analyzing only the brightest regions of the X-ray brightest clusters should mitigate most background-modeling concerns, it is not immediately clear how failing to account for OOT events might impact the results from pn data in these circumstances.}).
Our results from the individual detectors are consistent with this, albeit not at a statistically significant level in this case (right column in the figure).
Nevertheless, booyah.
The good agreement between Chandra and XMM furthermore suggests that our adoption of a monochromatic rather than an energy-dependent PSF model does not adversely effect the results, or at least not at a noticeable level compared with the intrinsic differences among the various instruments.

The impact of substituting the methodology presented in this work for our previous approaches, using the same spatial regions and temperature/metallicity binning,  is shown in Figure~\ref{fig:m1423_oldbg}.
For Chandra, the ``Blank sky + SFG'' results use the modified Cash statistic to account for a background estimated from the blank sky data, also including a single absorbed thermal emitter model for the SFG.
This additional SFG model was included even in earlier analysis, as the foreground is bright enough to be clearly detected in the cluster-free region \citep{Mantz1402.6212}.
The primary differences between these results and our baseline are a slight reduction in temperature and significant reduction in temperature precision at large radius, and an overestimate of the outermost density, all of which could easily be explained by limited statistics and/or a slight misnormalization of the blank sky spectra in the outskirts.
The figure also shows the outer two temperatures obtained when using the blank sky background without an SFG model (all other points are extremely similar to the ``Blank sky + SFG'' results); this choice straightforwardly leads to a significant underestimate of the outermost temperature.
For XMM, we compare our baseline results to those obtained using the spectrum of the cluster free region as a background estimate via the modified Cash statistic.
In this case, the outcome is a smaller increase in the outermost density, and a decrease in the outermost temperature, the latter likely being due to the disparate impact of OOT events originating from the cool core on the cluster spectra compared with the cluster-free spectra.

\subsection{Chandra: WARP\,J1415.1+3612}

WARP\,J1415 provides a test of the modeling for deep Chandra observations of a distant ($z=1.028$) cluster.
Like MACS\,J1423, WARP\,J1415 is relaxed and hosts a cool core; however, being at higher redshift, the surface brightness peak is less pronounced and the ICM signal becomes comparable to the background at smaller physical and angular radii.
WARP\,J1415 is also unique among known, massive clusters in having an extraordinary, super-solar Fe abundance in its core, measured in projection as $3.60^{+1.50}_{-0.85}$ in solar units within the central 12\,kpc \citep{Santos1111.3642}.
\citet{Mantz1706.01476} found a central, projected metallicity of $1.9\pm0.4$ Solar within a larger radius of 60\,kpc.
Our analysis will include deprojection, with the central metallicity constraint applying out to a comparable radius of 64.5\,kpc ($8''$ in projection).

\begin{table}
  \caption{Chandra observations of WARP\,1415}
  \hspace{-15mm}
  \centering
  \begin{tabular}{rccD}
    \hline
    \multicolumn{1}{c}{ObsID} & Det. & Date & \multicolumn{2}{c}{Clean exp.}\\
    &&& \multicolumn{2}{c}{(ks)}\\
    \hline
    \decimals
    4163 & ACIS-I & 2003-09-16 & 74.6 \\
    12256 & ACIS-S & 2010-08-28 & \hspace{1ex}118.5 \\
    12255 & ACIS-S & 2010-08-30 & 60.4 \\
    13118 & ACIS-S & 2010-09-01 & 44.6 \\
    13119 & ACIS-S & 2010-09-05 & 54.3 \\
    \hline
  \end{tabular}
  \label{tab:cl1415obs}
\end{table}

\begin{figure}
  \centering
  \includegraphics[scale=1.0]{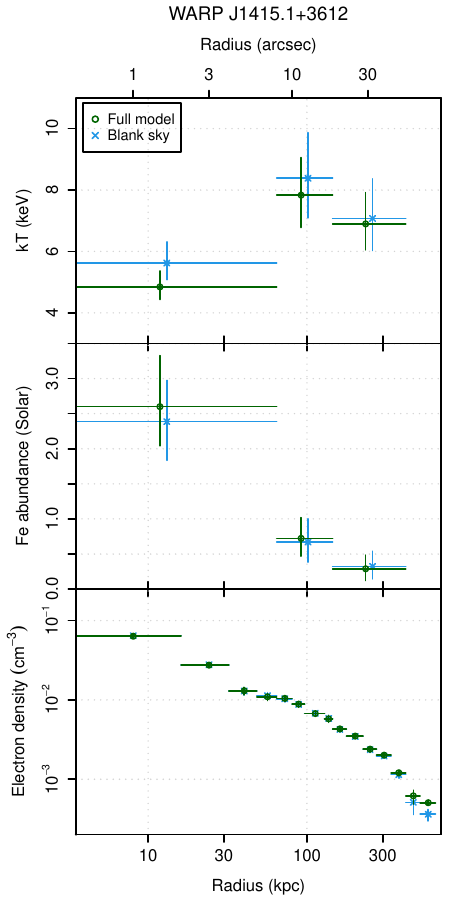}
  \caption{%
    Comparison of WARP\,J1415 profiles obtained from Chandra using the methods described in this work with blank-sky based approaches.
  }
  \label{fig:cl1415_oldbg}
\end{figure}

Figure~\ref{fig:cl1415_oldbg} shows the comparison of deprojected profiles using the methods described in this work with blank-sky background estimates.
Note that, unlike the last case, the SFG is not unusually bright in this field.
Our new analysis therefore uses the standard modeling priors based on ROSAT data, while the old-style analysis uses blank-sky estimates without an additional SFG component.
The two sets of constraints agree well generally, though the full modeling approach finds a slightly higher metallicity and lower temperature in the cluster center, as well as a slightly higher and better constrained density at large radii.
Given that the cluster center is substantially brighter than the average background (though, we note, significantly fainter than the center of MACS\,J1423), the differences in this region are presumably due to the sparsity of information in the corresponding blank-sky spectra and/or the binning required to use the modified Cash statistic (Section~\ref{sec:intro}).

More generally, our past work on the evolution of the ICM metallicity \citep{Mantz1706.01476} used spectra extracted in broad regions, modeled in projection rather than in 3D, precisely because binning in energy space inevitably reduces the accuracy with which emission lines can be measured.
This compromise comes at the expense of being able to constrain the deprojected density profile at high spatial resolution simultaneously with the metallicity profile.
With forward modeling of all foreground and background components, the classical Cash statistic can be used without binning, avoiding this shortcoming.

\subsection{Chandra: MACS J0717.5+3745}

We now turn to cluster MACS\,J0717 ($z=0.546$), a complex merger where previous works have reported projected ICM temperatures from $\sim10$--25\,keV based on Chandra data directly \citep{Ma0901.4783, Mroczkowski1205.0052, Sayers1312.3680} and the combination of Chandra brightness with measurements of the thermal Sunyaev-Zel'dovich effect \citep{Adam1706.10230}.
Measuring temperatures of the hot ICM when the energy $kT/(1+z)$ falls beyond the instrument's bandpass is inherently challenging, and, one might expect, particularly sensitive to the spectral modeling of the background.
Because this cluster is far from spherically symmetric, we forego our usual deprojected analysis and instead follow the above-cited authors in constraining a map of the projected temperature.
This has the added feature of removing the correlations between measured temperatures that result from fitting a deprojected model, making it, in this sense, a cleaner test of the impact of the background modeling.
(Correlations due to linking the metallicity and the appropriate parameters of the background models are still present.)
The data analyzed in this section are summarized in Table~\ref{tab:macs0717obs}.

\begin{table}
  \caption{Chandra observations of MACS\,J0717}
  \hspace{-15mm}
  \centering
  \begin{tabular}{rccc}
    \hline
    \multicolumn{1}{c}{ObsID} & Det. & Date & \multicolumn{1}{c}{Clean exp.}\\
    &&& \multicolumn{1}{c}{(ks)}\\
    \hline
    \decimals
    1655 & ACIS-I & 2001-01-29 & 16.5 \\
    4200 & ACIS-I & 2003-01-08 & 59.2 \\
    16305 & ACIS-I & 2013-12-11 & 76.0 \\
    16235 & ACIS-I & 2013-12-13 & 66.8 \\
    \hline
  \end{tabular}
  \label{tab:macs0717obs}
\end{table}

\begin{figure}
  \centering
  MACS J0717.5+3745\smallskip\\
  \includegraphics[scale=0.35]{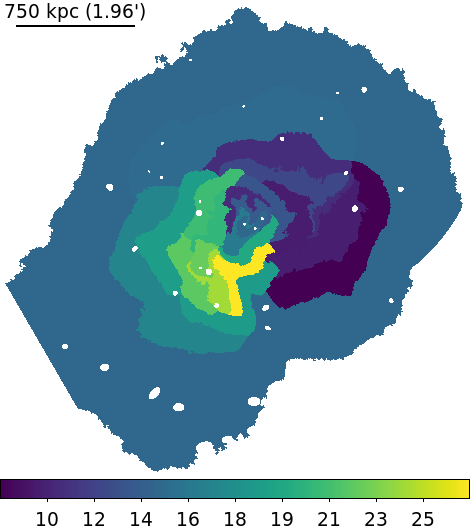}\\
  \includegraphics[scale=0.35]{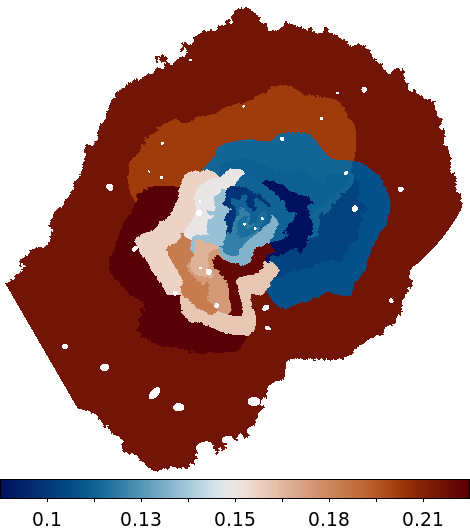}\\
  \includegraphics[scale=0.35]{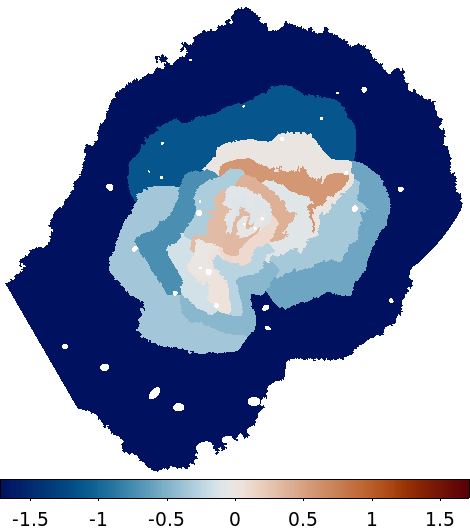}
  \caption{%
    Top: Best-fitting projected temperature map of MACS\,J0717 using the background modeling techniques from this work.
    Center: Symmetrized fractional uncertainty in those best-fitting temperatures.
    Bottom: Change in best-fitting temperature when using a blank sky background rather than modeling, as a fraction of the latter's uncertainty.
    All scale bars have units of keV. 
  }
  \label{fig:m0717_maps}
\end{figure}

 To define regions for spectral analysis, we applied the {\sc contbin} algorithm \citep{Sanders0606528} to 0.6--7.0\,keV band stacked images of the cluster, binned to 0.984$''$ resolution, with a target signal-to-noise of 55 and a geometric constraint parameter of 2.
These regions were manually split in some cases to eliminate ``gerrymandering'', and our final analysis links temperatures across several of the faintest, similar-brightness regions where they would not otherwise be constrained, forming a quasi-annular region at the largest radii included.
We assume a common metallicity among all regions, allowing the ICM temperatures and normalizations to fit independently, and utilizing cluster-free regions to constrain the background model.

Figure~\ref{fig:m0717_maps} shows the best-fitting map of projected ICM temperature and its fractional uncertainty from our analysis (left and center panels, respectively).
We find a similar range and qualitative distribution of temperatures to previous work \citep{Ma0901.4783, Adam1706.10230}.
The bottom panel of the figure shows the difference between the best-fitting temperatures obtained using blank sky backgrounds and those from our baseline analysis, in units of the uncertainty from the baseline analysis.
We find compatible temperature constraints between the methods (within $1\sigma$) throughout the brightest part of the cluster.
The only significant deviations occur in the quasi-annular region at the largest radii, where the cluster signal is faintest.
Note that we apply our standard process of restricting the energy range of the spectra such that the background is never truly dominant in both approaches, which should reduce the impact of background modeling choices compared with using the full bandpass.

\subsection{Chandra+XMM: SPT-CL J0459$-$4947} \label{sec:spt0459}

We conclude with an example of a high-redshift ($z=1.71$) cluster analyzed with both Chandra and XMM observations, SPT-CL\,J0459.
This is more typical of recent observations targeting the most distant clusters discovered, whose high redshift, combined with the ongoing reduction in Chandra's soft response, makes both observatories necessary to obtain a complete picture.
In this case, the Chandra data (122\,ks clean exposure) are significantly less sensitive than the XMM observations (377\,ks MOS/322\,ks pn clean).
Table~\ref{tab:spt0459obs} lists the data for this system.

\begin{table}
  \caption{Observations of SPT\,J0459}
  \hspace{-15mm}
  \begin{tabular}{crccD}
    \hline
    Obs. & \multicolumn{1}{c}{ObsID} & Det. & Date & \multicolumn{2}{c}{Clean exp.}\\
    &&&& \multicolumn{2}{c}{(ks)}\\
    \hline
    \decimals
    CXO & 17501 & ACIS-I & 2015-11-19 & 22.5 \\
    CXO & 18711 & ACIS-I & 2015-11-28 & 22.4 \\
    CXO & 17502 & ACIS-I & 2016-04-14 & 12.0 \\
    CXO & 18824 & ACIS-I & 2016-04-17 & 21.7 \\
    CXO & 17211 & ACIS-I & 2016-05-21 & 13.6 \\
    CXO & 18853 & ACIS-I & 2016-05-22 & 30.2 \\
    XMM & 0801950501 & MOS1 & 2017-04-14 & 18.0 \\
    \ldots & \ldots & MOS2 & \ldots & 18.6 \\
    \ldots & \ldots & pn & \ldots & 14.4 \\
    XMM & 0801950101 & MOS1 & 2017-07-28 &  \hspace{1ex}100.8 \\
    \ldots & \ldots & MOS2 & \ldots & 102.9 \\
    \ldots & \ldots & pn & \ldots & 88.6 \\
    XMM & 0801950201 & MOS1 & 2017-07-30 & 96.2 \\
    \ldots & \ldots & MOS2 & \ldots & 95.8 \\
    \ldots & \ldots & pn & \ldots & 85.1 \\
    XMM & 0801950301 & MOS1 & 2017-10-20 & 94.7 \\
    \ldots & \ldots & MOS2 & \ldots & 94.3 \\
    \ldots & \ldots & pn & \ldots & 81.8 \\
    XMM & 0801950401 & MOS1 & 2017-11-07 & 66.7 \\
    \ldots & \ldots & MOS2 & \ldots & 65.7 \\
    \ldots & \ldots & pn & \ldots & 51.7 \\
    \hline
  \end{tabular}
  \label{tab:spt0459obs}
\end{table}

Previous analyses of this cluster, constraining the gas mass profile using relatively finely spaced annular regions, were unable to simultaneously constrain more than a single temperature and metallicity.
Instead, to avoid the impact of binning spectra in order to use the modified Cash statistic, 2-temperature and metallicity profiles were constrained using much larger annuli \citep{Mantz2006.02009, Flores2108.12051}.
Here we show deprojected results obtained from a finer set of annuli, with multiple temperatures and metallicities free, using the modified Cash statistic approach and an empirical backgrounds extracted from either radii $3'$--$12'$ or $3'$--$5'$ from the cluster center (the latter is the background region adopted in the works cited above), and compare them with those obtained using forward modeling of the background.
We follow the above-cited works by directly employing only the XMM data in this deprojection, with the Chandra data used to define the point-source masks and model the residual CXB contamination.

\begin{figure}
  \centering
  \includegraphics[scale=1.0]{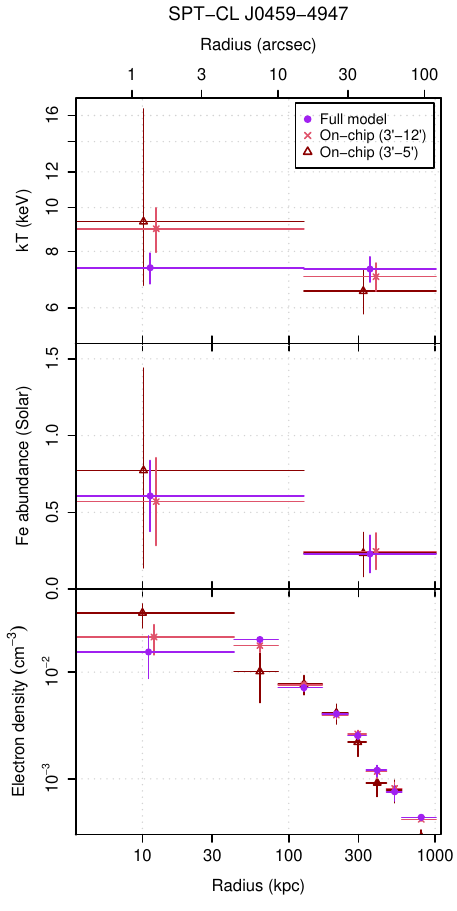}
  \caption{%
    Comparison of SPT\,0459 profiles obtained from XMM using the methods described in this work with a background estimate directly from the science observation.
    Note that the temperature axis is logarithmic in this case, to accommodate the large uncertainty in the central value when using the $3'$--$5'$ on-chip background.
  }
  \label{fig:spt0459}
\end{figure}

Figure~\ref{fig:spt0459} compares profiles obtained using our methods with those using each of the cluster-free regions of the science observation to estimate the total background.
Interestingly, the use of the larger on-chip background region significantly improves the temperature and metallicity precision at small radii compared with the smaller on-chip region, although the resulting temperature constraint is marginally inconsistent with what we obtain from the full model.
This bias may be due to differences in the QPB between the central and outer parts of the detector, although we have not investigated further.
Apart from this central temperature and the central density, the results using the forward modelled backgrounds and the classical Cash statistic are consistent with the others, and provide higher precision.
Similarly to, though much more dramatically than, the Chandra-only analysis of WARP\,J1415, we see that the difference in constraining power is more significant at small radii than large radii.
The enhanced impact of the modeling in this case may be due to the significant PSF mixing between small and large radii exacerbating the challenge of extracting information about the cluster center when using empirical background estimation and/or binned spectra.
In future work, we will test whether these features are common to high-redshift clusters probed with this style of joint XMM+Chandra analysis.

\section{Summary and Conclusion} \label{sec:conclusion}

We have variously synthesized and developed procedures for forward modeling the foregrounds and quiescent backgrounds affecting Chandra and XMM observations.
Although descriptions of the particle-induced background and OOT events remain ultimately empirical, this framework nevertheless provides a generative model that can be applied without the requirement of binning any spectra being analyzed.
We demonstrate the application of these methods to the analysis of several galaxy clusters spanning a range of physical properties and redshifts, comparing old and new methodologies.
Both approaches generally seek to be conservative in the treatment of the background (i.e.\ limit the impact of residual systematic uncertainties), and as expected the results are consequently largely consistent with one another.
However, we do see suggestions in each case that the fully forward-modeled approach enables more information to be reliably extracted from the data, as it should.
Expanding the comparison to a larger number of systems observed in various conditions will prove illuminating.
We have not addressed the issue of residual contamination from soft protons in this work, as it was unnecessary for the data considered here, but modeling it would be a natural next addition.

Adopting generative models for the various X-ray backgrounds allows inferences to be drawn on a statistically firm and transparent footing.
Rather than asking whether the background spectrum extracted from a given observation and region is appropriate, or how to best propagate the ``error'' from subtracting it, we can instead debate the more scientifically oriented questions of whether the model and its priors are physically defensible and adequate to explain the data.
The approach alone is not a panacea, as the backgrounds still present a fundamental limitation on what we can learn from the data.
In particular, once the source of interest becomes fainter than Poisson scatter in the background realization in a given region and/or energy range, no methodology will be able to detect or constrain it with high significance.

Looking forward, the successors to Chandra (AXIS, if approved) and XMM (NewAthena) will both be deployed to relatively high-particle-background environments, making background mitigation critical to the study of faint, diffuse targets with either.
While the methods employed in this work benefit from decades of study of the X-ray foreground and backgrounds, appropriately designed calibration efforts can make them applicable to future missions almost immediately.
We particularly note the importance of filter-closed or equivalent data in order to constrain the in-situ QPB as well as regular monitoring of both hard and soft calibration lines to constrain changes in the effective area.

\begin{acknowledgments}
  We thank Hiromasa Suzuki for discussions about his Chandra QPB model, \'Akos Bogd\'an for discussions about ACIS calibration, and all the researchers who have spent their professional lives studying the various X-ray backgrounds, so that the rest of us wouldn't need to.
  
  This research has made use of data obtained from the Chandra Data Archive provided by the Chandra X-ray Center (CXC)
  and
  observations obtained with XMM-{\it Newton}, an ESA science mission with instruments and contributions directly funded by ESA Member States and NASA.
  Support for this work was provided in part by the National Aeronautics and Space Administration through Chandra Award Numbers GO2-23113A, GO3-24102X, GO3-24111A, GO3-24113X, GO4-25086X and GO5-26096X issued by the Chandra X-ray Center, which is operated by the Smithsonian Astrophysical Observatory for and on behalf of the National Aeronautics Space Administration under contract NAS8-03060.
  We acknowledge support from the National Aeronautics and Space Administration under Grant Numbers 80NSSC21K0759, 80NSSC24K0330, 80NSSC23K1579 and 80NSSC23K1577, issued through the XMM-Newton Guest Observer Facility,
  as well as through Astrophysics Data Analysis award 80NSSC24K1401.
  We acknowledge support from the U.S. Department of Energy under contract number DE-AC02-76SF00515.
\end{acknowledgments}

%

%
\facilities{CXO, ROSAT, XMM}

\software{
  ACX (\citealt{Smith2012AN....333..301S, Smith1406.2037}; \url{http://atomdb.org/CX}),
  Astropy \citep{astropy1304.002, Astropy-Collaboration1307.6212, Astropy-Collaboration1801.02634, Astropy-Collaboration2206.14220},
  CIAO \citep{Fruscione2006SPIE.6270E..1VF, ciao1311.006},
  contbin \citep{Sanders0606528, contbin1609.023},
  ESAS (\url{https://heasarc.gsfc.nasa.gov/docs/xmm/esas}),
  HEASOFT \citep{heasoft1408.004},
  LMC \citep{lmc1706.005},
  MARX \citep{Davis2012SPIE.8443E..1AD, marx1302.001},
  SAS \citep{sas1404.004},
  SXRBG \citep{sxrbg1904.001},
  XSPEC \citep{Arnaud1996ASPC..101...17A, xspec9910.005}
}

\appendix

\section{ACIS Calibration Correction} \label{sec:cal_correct}

\begin{figure*}[ht!]
    \centering
    \includegraphics[scale=0.25]{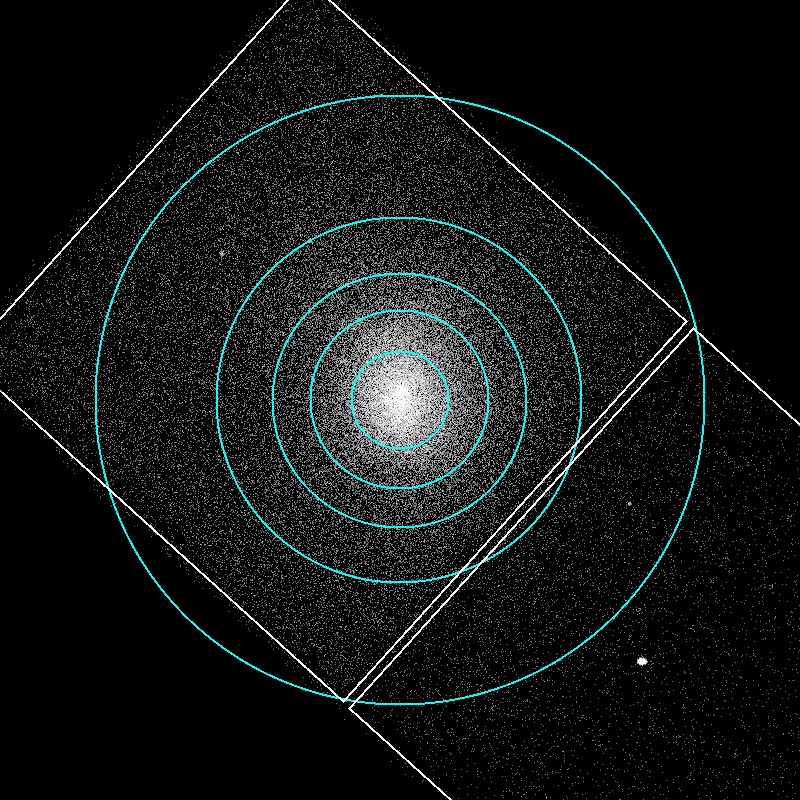}
    \hspace{5mm}
    \includegraphics[scale=0.25]{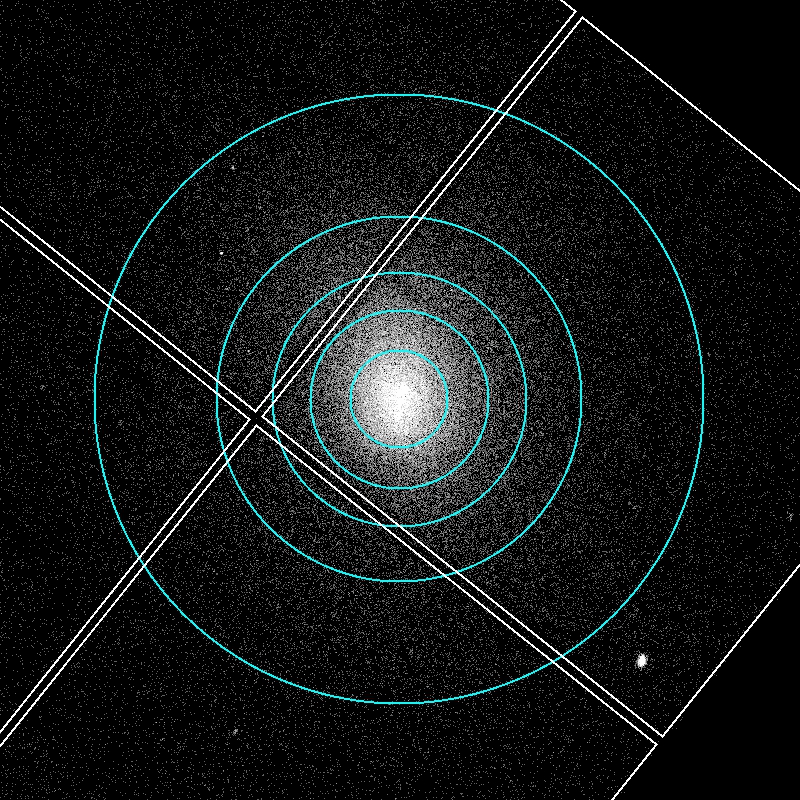}
    \caption{%
    Example images of A1795 observed with ACIS-S (left) and ACIS-I (right), showing the 5 annular regions used in our investigation of the calibration.
    We employ only those observations where the cluster center is positioned near the nominal aimpoint, although the positioning is not identical in all cases.
    }
    \label{fig:cal_images}
\end{figure*}

While performing a preliminary analysis of the Chandra data for SPT\,J2215$-$3537, one of the first applications of this work \citep{Stueber2601.14425}, we noticed a systematic offset between temperatures measured from the observations associated with the 2 proposals targeting this cluster.\footnote{\citet{Watson2511.00250} have noted the same issue in ACIS observations of A2029 spanning 2000--2023, and suggest a correction, notionally similar to ours, based on those data.}
These differ both in the date of observation and in the use of ACIS-I (observed in Aug.\ 2020) vs ACIS-S (observations spread between Dec.\ 2021 and Aug.\ 2023).
To investigate whether this was a reproducible effect and its potential causes, we turned to observations of A1795, which span the duration of the Chandra mission and a range of focal plane temperatures.
Specifically, our analysis uses only ``on-axis'' observations, where the cluster center is at or near the aimpoint on either chip I3 or S3, as opposed to the ``offset'' pointings that are among the regular calibration observations made since 2010.
We further exclude ObsIDs 494, the only warm-focal-plane, on-axis observation from early in the mission, and 21839, which is excessively flared.
The observations are summarized in Table~\ref{tab:a1795obs}.

\begin{table}
  \caption{Chandra observations of A1795}
  \hspace{-15mm}
  \centering
  \begin{tabular}{rcccDrcccD}
    \hline
    \multicolumn{1}{c}{ObsID} & Det. & Date & FP Temp. & \multicolumn{2}{c}{Clean exp.} & \multicolumn{1}{c}{ObsID} & Det. & Date & FP Temp. & \multicolumn{2}{c}{Clean exp.}\\
    &&& ($^\circ$C) & \multicolumn{2}{c}{(ks)} & &&& ($^\circ$C) & \multicolumn{2}{c}{(ks)}\\
    \hline
    \decimals
    493    &  ACIS-S  &  2000-03-21  &  $-119.7$  &  \hspace{2ex}19.6  &         20651  &  ACIS-S  &  2018-04-12  &  $-113.2$  &  \hspace{2ex}14.5  \\  
    3666   &  ACIS-S  &  2002-06-10  &  $-119.7$  &  13.3  &                     20644  &  ACIS-I  &  2018-04-13  &  $-118.1$  &  18.8  \\  
    5287   &  ACIS-S  &  2004-01-14  &  $-119.5$  &  14.3  &                     20642  &  ACIS-S  &  2018-11-16  &  $-115.5$  &  14.9  \\  
    5289   &  ACIS-I  &  2004-01-18  &  $-119.7$  &  14.1  &                     20643  &  ACIS-I  &  2018-11-25  &  $-119.5$  &  19.5  \\  
    6160   &  ACIS-S  &  2005-03-20  &  $-119.2$  &  14.8  &                     21832  &  ACIS-I  &  2019-04-08  &  $-119.5$  &  19.2  \\  
    6162   &  ACIS-I  &  2005-03-28  &  $-119.1$  &  13.3  &                     21831  &  ACIS-I  &  2019-11-26  &  $-114.4$  &  19.8  \\  
    10900  &  ACIS-S  &  2009-04-20  &  $-119.7$  &  15.8  &                     21830  &  ACIS-S  &  2019-11-29  &  $-117.3$  &  14.9  \\  
    12028  &  ACIS-S  &  2010-05-10  &  $-119.7$  &  15.0  &                     22831  &  ACIS-I  &  2020-04-13  &  $-119.1$  &  18.1  \\  
    12026  &  ACIS-I  &  2010-05-11  &  $-119.7$  &  14.9  &                     22838  &  ACIS-S  &  2020-04-15  &  $-113.2$  &  13.9  \\  
    13108  &  ACIS-I  &  2011-03-10  &  $-119.4$  &  14.6  &                     22829  &  ACIS-S  &  2020-12-04  &  $-119.7$  &  14.9  \\  
    13106  &  ACIS-S  &  2011-04-01  &  $-119.7$  &  9.9   &                     22830  &  ACIS-I  &  2020-12-04  &  $-117.6$  &  19.5  \\  
    14270  &  ACIS-I  &  2012-03-25  &  $-119.7$  &  14.3  &                     24602  &  ACIS-I  &  2021-04-22  &  $-118.4$  &  20.0  \\  
    14268  &  ACIS-S  &  2012-03-26  &  $-118.2$  &  9.9   &                     24609  &  ACIS-S  &  2021-04-26  &  $-114.2$  &  14.9  \\  
    15485  &  ACIS-S  &  2013-04-21  &  $-119.7$  &  9.9   &                     24600  &  ACIS-S  &  2021-11-15  &  $-114.4$  &  14.9  \\  
    15487  &  ACIS-I  &  2013-06-02  &  $-119.4$  &  12.8  &                     24601  &  ACIS-I  &  2021-11-20  &  $-115.3$  &  19.0  \\  
    16432  &  ACIS-S  &  2014-04-02  &  $-119.7$  &  9.9   &                     25670  &  ACIS-I  &  2022-04-19  &  $-115.7$  &  18.6  \\  
    16434  &  ACIS-I  &  2014-04-02  &  $-119.4$  &  14.6  &                     25677  &  ACIS-S  &  2022-04-29  &  $-115.3$  &  14.9  \\  
    17399  &  ACIS-I  &  2015-04-07  &  $-119.5$  &  14.4  &                     25668  &  ACIS-S  &  2022-12-22  &  $-112.4$  &  13.8  \\  
    17397  &  ACIS-S  &  2015-04-16  &  $-114.2$  &  9.9   &                     25669  &  ACIS-I  &  2022-12-25  &  $-111.6$  &  18.1  \\  
    17683  &  ACIS-I  &  2015-07-20  &  $-119.7$  &  13.1  &                     27027  &  ACIS-S  &  2023-04-02  &  $-116.2$  &  15.4  \\  
    17685  &  ACIS-S  &  2015-07-20  &  $-115.5$  &  9.9   &                     27020  &  ACIS-I  &  2023-04-23  &  $-116.3$  &  8.7   \\  
    18424  &  ACIS-I  &  2015-12-03  &  $-116.2$  &  15.1  &                     27806  &  ACIS-I  &  2023-04-23  &  $-117.3$  &  8.9   \\  
    18423  &  ACIS-S  &  2015-12-06  &  $-116.3$  &  9.9   &                     27018  &  ACIS-S  &  2024-01-09  &  $-112.1$  &  16.3  \\  
    18425  &  ACIS-S  &  2016-04-08  &  $-118.6$  &  9.9   &                     27019  &  ACIS-I  &  2024-01-13  &  $-112.8$  &  16.8  \\  
    18427  &  ACIS-I  &  2016-04-08  &  $-117.3$  &  14.6  &                     28471  &  ACIS-S  &  2024-06-07  &  $-112.8$  &  15.3  \\  
    19868  &  ACIS-S  &  2016-11-23  &  $-117.0$  &  15.9  &                     28464  &  ACIS-I  &  2024-06-27  &  $-119.7$  &  17.5  \\  
    19869  &  ACIS-I  &  2016-11-25  &  $-119.7$  &  19.8  &                     29623  &  ACIS-I  &  2025-02-05  &  $-113.7$  &  16.3  \\  
    19870  &  ACIS-I  &  2017-04-08  &  $-114.7$  &  19.3  &                     29624  &  ACIS-S  &  2025-02-29  &  $-112.0$  &  14.9  \\  
    19877  &  ACIS-S  &  2017-04-23  &  $-119.4$  &  14.9  &                    29613  &  ACIS-I  &  2025-06-05  &  $-117.8$  &  16.6  \\  
    19968  &  ACIS-S  &  2017-11-19  &  $-119.4$  &  14.9  &                    29620  &  ACIS-S  &  2025-06-24  &  $-113.7$  &  13.9 \\   
    19969  &  ACIS-I  &  2017-12-01  &  $-115.5$  &  19.3  \\
    \hline
  \end{tabular}
  \label{tab:a1795obs}
\end{table}

\begin{figure*}[ht!]
    \centering
    \includegraphics[scale=1]{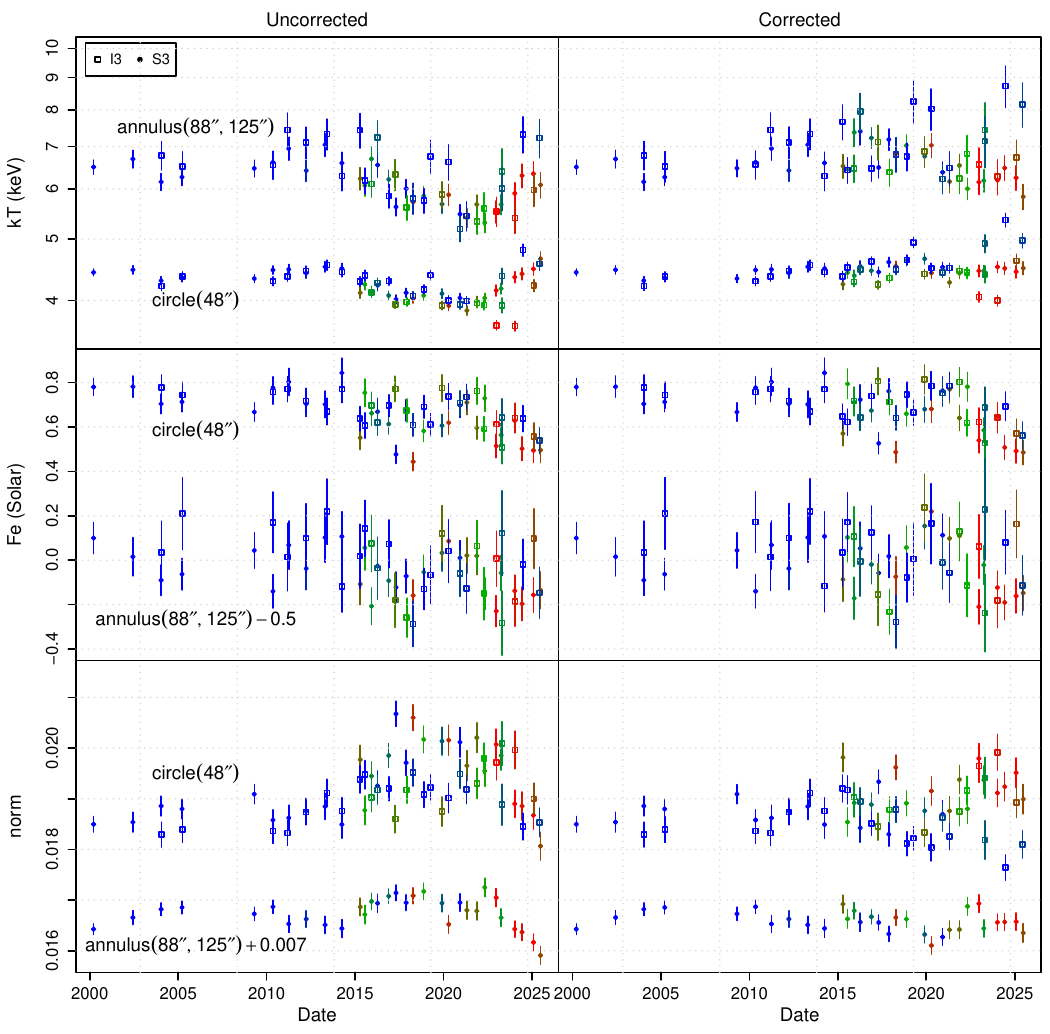}
    \caption{%
    {\sc apec} parameters measured from on-axis observations of A1795 over the course of the Chandra mission.
    The left and right columns, respectively, show results without and with the correction described in the text.
    Points are color coded according to the focal plane temperatures stored in the event file headers, from $-119.6\,^{\circ}$C (blue) to $-111.5\,^{\circ}$C (red).
    Some quantities are offset by a constant, as indicated, for clarity.
    Note that the legend refers to which CCD the cluster center appears on;
    in particular, the ``I3'' data in the annulus reflect a joint fit including data from both CCDs I2 and I3 (normalizations are not shown for this case because there are gaps in the coverage of the annulus).
    }
    \label{fig:cal_correction}
\end{figure*}

Our investigation uses the set of annuli shown in Figure~\ref{fig:cal_images}, centered on $13^\mathrm{h}48^\mathrm{m}52\overset{\mathrm{s}}{.}7$+$26^\circ35'27''$ (J2000), the target position used in the regular calibration observations.
The radii defining these annuli are: $48''$, $88''$, $125''$, $180''$ and $300''$.
The circular region of radius $48''$ was chosen to be as large as possible while still being fully imaged, accounting for our standard masking of chip edges, within chip I3 in every ACIS-I observation; similarly, all regions with radii $\leq125''$ are fully observed on chip S3 in every ACIS-S observation.
Otherwise, the radii are chosen to provide broadly similar constraining power, without enclosing too large a range of temperatures.
The outer 3 annuli lie in the roughly isothermal part of the temperature profile according to previous work (e.g.\ \citealt{Snowden0710.2241}), as our measurements verify.

\begin{figure*}
    \centering
    \includegraphics[scale=1]{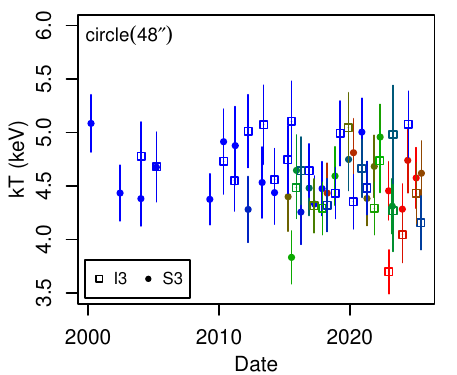}
    \hspace{5mm}
    \includegraphics[scale=1]{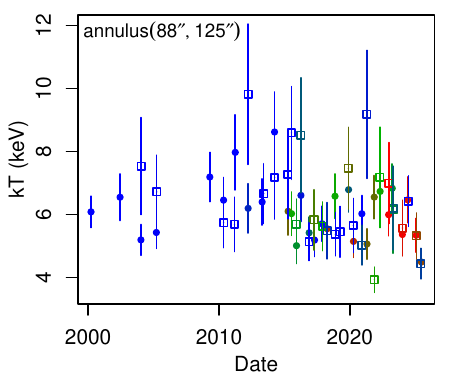}
    \caption{%
      Temperatures measured for A1795, presented as in Figure~\ref{fig:cal_correction}, for the central region (left) and an annulus at intermediate radii (right).
      Energies $<3$\,keV are ignored in these fits, at the expense of precision; however,  no time-dependent correction appears necessary.
    }
    \label{fig:cal_highE}
\end{figure*}

We employed the full forward modeling of foregrounds and backgrounds, as described above, but have verified that our results in this section change negligibly if we instead follow the standard practice of using a blank sky background, normalized according to the high-energy count rate.
For simplicity, our model of the cluster emission in this case is a simple absorbed thermal emission model ({\sc phabs*apec}) without deprojection.
The results shown in this section use {\sc caldb} 4.12.2 and {\sc ciao} 4.17, the latest versions at this writing, although our original investigation used an earlier version; our results are qualitatively consistent across versions since 4.11.0.%
\footnote{While this paper was in revision, {\sc ciao 4.18} and {\sc caldb} 4.12.3 were released. This update provides a clear improvement over the uncorrected results shown in this section, although smaller time-dependent trends and the significant scatter in ACIS-I measurements remain. Table~\ref{tab:calcor} provides our best correction for this newer calibration version as well.}

\begin{figure*}
    \centering
    \includegraphics[scale=0.8]{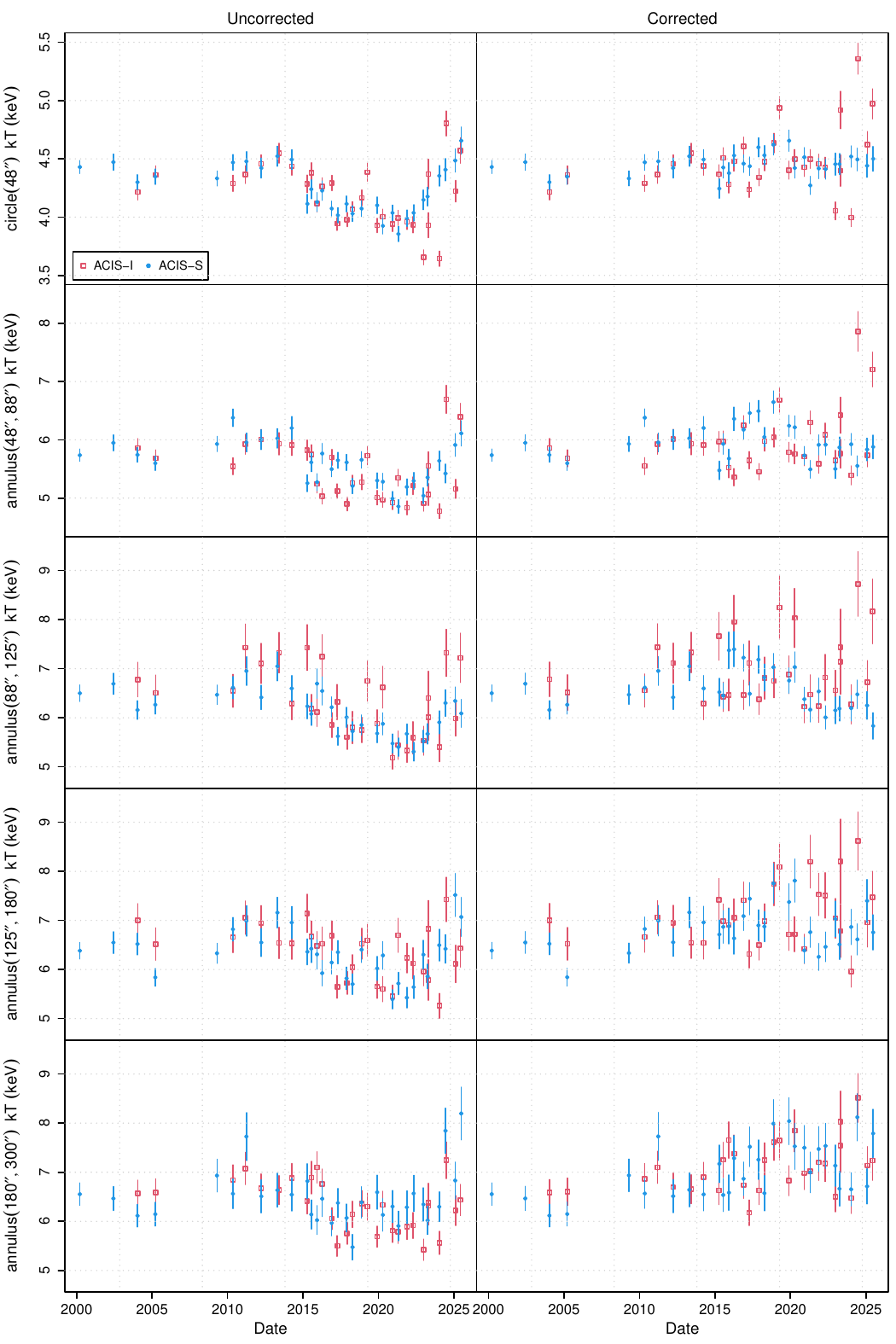}
    \caption{%
    Comparison of uncorrected and corrected temperature measurements of A1795, as in Figure~\ref{fig:cal_correction}, for 5 regions at different cluster radii.
    }
    \label{fig:cal_allradii}
\end{figure*}

The left column of Figure~\ref{fig:cal_correction} shows constraints on the {\sc apec} temperature, abundance and normalization parameters from each observation in the central circle and an intermediate annulus; temperature measurements for all regions are shown in Figure~\ref{fig:cal_allradii}.
We see relatively stable behavior prior to 2015, and thereafter a noticeable reduction in temperature and metallicity, and an increase in normalization (the latter more pronounced in the central circular region).
The reversal of these trends since 2022 is a product of the gain update in {\sc caldb} 4.12.2, and is largely corrected in 4.12.3 (see Footnote~20).
The ACIS-I observations appear to show greater scatter in these late observations compared with ACIS-S, while neither displays an obvious residual trend with focal plane temperature (color coding in the figure).
We note that these trends largely vanish when excluding the lowest energies from the fit rather than using the full 0.6--7.0\,keV band; however, it is necessary to remove all the data at energies $<3$\,keV to convincingly accomplish this (Figure~\ref{fig:cal_highE}).

The presence of emission from gas over a range in temperature, whether due to projection or genuine multiphase gas in the cluster core, potentially complicates the interpretation of these measurements.
However, we see no indication of a significant impact from multiphase gas.
Generically, one would expect to see a discrepancy between single-temperature fits using the front-illuminated and back-illuminated CCDs (due to their different soft responses) in this case, with these differences varying with time and the region considered (i.e.\ the potentially multiphase core vs the more isothermal larger radii).
Instead, we see good agreement between front- and back-illuminated CCDs across regions and as a function of time, even as both depart from being constant in time.

\begin{figure*}
    \centering
    \includegraphics[scale=1]{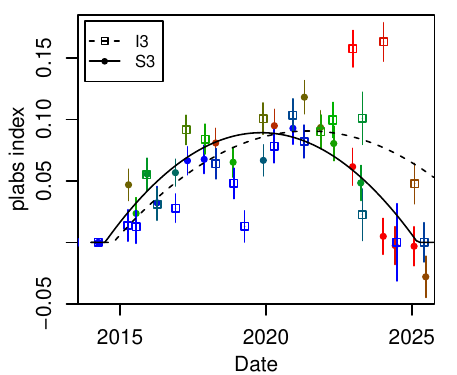}
    \hspace{5mm}
    \includegraphics[scale=1]{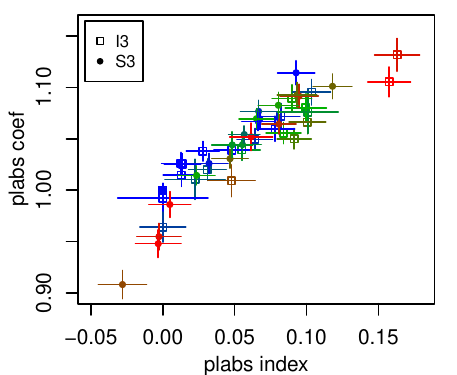}
    \caption{%
    Left: Time dependence of the {\sc plabs} index parameter fitted to the ACIS-I and ACIS-S data for A1795 separately, while linking the {\sc apec} model parameters.
    This analysis assumes that pre-2015 data require no correction.
    Right: The relationship between the two {\sc plabs} parameters in the same fits.
    As in Figure~\ref{fig:cal_correction}, colors reflect the focal plane temperature, and the ACIS-I data generally display larger scatter.
    }
    \label{fig:cal_plabs}
\end{figure*}

Given this consistency, our strategy is to use the central region that is fully covered in every observation to derive a correction, which is then tested using the data at larger radii. We have verified that essentially identical results are obtained for ACIS-S when using the $88''$--$125''$ annulus to determine a correction that is applied to the other regions.
Specifically, we investigated whether a simple multiplicative component in the {\sc xspec} model fitted to the data, with time dependent parameters, could correct the observed trends, assuming that pre-2015 data require no correction.
To accomplish this, we linked the {\sc apec} parameters across ObsIDs, allowing the parameters of the new model to be free for post-2015 observations.
We did, however, analyze the ACIS-I and ACIS-S data independently.
We investigated using either an absorption edge ({\sc edge}) or power-law absorption ({\sc plabs}) model, finding that {\sc plabs} was more effective at removing the trends with time.
This model simply multiplies the predicted spectrum by a power law, $A(E/\mathrm{keV})^\alpha$, with the coefficient, $A$, and index, $\alpha$, being free parameters.
The left panel of Figure~\ref{fig:cal_plabs} shows the time dependence of the index that results from requiring consistent {\sc apec} parameters across the data in the central circle.
The two {\sc plabs} parameters are strongly degenerate, and display a mild residual dependence on the focal plane temperature, as seen in the right panel of the figure.

The correction we devise is as follows, and is arrived at independently for ACIS-I and ACIS-S.
First, linear models relating the {\sc plabs} coefficient to the index and focal plane temperature were fit to the data, with the constraint that the models reduce to no correction (index zero and coefficient unity) at a focal plane temperature of $-119.6\,^{\circ}$C.
Second, we fit quadratic models to describe the time dependence of the {\sc plabs} index since 2015.
With these two ingredients, the date and focal plane temperature of an observation can be used to define an energy dependent correction to apply in spectral modeling.%
\footnote{In the absence of the mid-2024 and later data, which were available only for our analysis of {\sc caldb} 4.12.2, the large scatter of the ACIS-I measurements made it much more difficult to discern trends. For the earlier {\sc caldb} versions for which we provide corrections in Table~\ref{tab:calcor}, the procedure described here was therefore simplified slightly. Specifically, we fitted the linear model relating the {\sc plabs} coefficient and index to only the ACIS-S data, applying it to both detectors, and used a linear rather than quadratic model for the time dependence of the index for ACIS-I.}

The right columns of Figures~\ref{fig:cal_correction} and \ref{fig:cal_allradii} show the measured {\sc apec} parameters for different regions in A1795 when the correction defined by this procedure is applied; we see that the trends have largely been removed, even if (as expected) some residual scatter remains, particularly in the ACIS-I data.
Significantly, this is the case not just in the cluster center where the correction is defined, but also for the annuli at larger radii, extending into the isothermal part of the temperature profile.
The exception is the outermost annulus ($180''$--$300''$), where the correction appears to be too strong, resulting in a slightly increasing trend of temperature with time.
This suggests a possible relationship with the correction for contamination on the optical blocking filter, which is expected to spatially vary along with the temperature of the filter itself.
In particular, this pattern is expected to be relatively simple within a few arcmin of the nominal aimpoints on ACIS-I and -S, and vary more quickly at larger radii on ACIS-I and within $\sim1'$ of the edges of ACIS-S adjacent and opposite to ACIS-I (\citealt{Vikhlinin.CXO.OBF.memo}; cf Figure~\ref{fig:cal_images}).
The offset calibration pointings could be used to investigate this further, though we have not done so at this time.

Table~\ref{tab:calcor} shows the parameters that define the correction for several recent {\sc caldb} versions.
To be explicit, we have implemented these as follows.
The index of the power-law correction is found as
\begin{equation}
\alpha = a_0 + a_1 \Delta t + a_2 \Delta t^2
\end{equation}%
where $\Delta t$ is the difference between the observation time and 2020-01-01 in days.
The {\sc plabs} coefficient is
\begin{equation}
A = 1 + c_\alpha\alpha + c_T \Delta T,
\end{equation}%
where $\Delta T$ is the difference between the focal plane temperature and $-119.554^\circ$C in $^\circ$C.
If the above formulae would yield $\alpha<0$ for dates prior to 2020, we set $\alpha=0$ and $A=1$ (i.e., no correction at early times; see Figure~\ref{fig:cal_plabs}).
The correction can either be applied using the {\sc plabs} model with parameters $\alpha$ and $A$ in {\sc xspec}, or equivalently by directly multiplying the effective area curve by the function $A(E/1\,\mathrm{keV})^\alpha$.

\begin{table*}
  \caption{Calibration Correction Parameters}
  \hspace{-2.5cm}
  \begin{tabular}{c@{\hspace{5ex}}ccccc@{\hspace{7ex}}ccccc}
    \hline
    & \multicolumn{5}{c}{ACIS-I} & \multicolumn{5}{c}{ACIS-S} \\
    {\sc caldb} & $a_0$ & $a_1$ & $a_2$ & $c_\alpha$ & $c_T$ & $a_0$ & $a_1$ & $a_2$ & $c_\alpha$ & $c_T$ \\
    & $(10^{-2})$ & $(10^{-5})$  & $(10^{-8})$ & $(10^0)$ & $(10^{-3})$ & $(10^{-2})$ & $(10^{-5})$ & $(10^{-8})$ & $(10^0)$ & $(10^{-3})$ \\
    \hline
    4.11.0 & 7.440 & 3.440 & 0.0 & 1.188242 & $-4.0804$ & 8.152 & \hspace{2ex}0.2716 & $-1.718$ & 1.188242 & $-4.0804$ \\
    4.11.1 & 7.440 & 3.440 & 0.0 & 1.188404 & $-5.3303$ & 7.892 & \hspace{2ex}1.1940 & $-1.189$ & 1.188404 & $-5.3303$ \\
    4.12.0 & 8.142 & 3.838 & 0.0 & 1.188404 & $-5.3303$ & 7.892 & \hspace{2ex}1.1940 & $-1.189$ & 1.188404 & $-5.3303$ \\
    4.12.2 & 8.634 & 1.610 & $-1.524$ & 1.014497   & $-6.2150$   & 8.912 & $-0.4829$ & $-2.548$ & 1.204025 & $-4.6903$ \\
    4.12.3 & 4.751 & 2.602 & 0.0 &  1.142516  & $-7.3920$   & 3.042 & \hspace{2ex}1.6660 & 0.0 & 0.9891564 & $-2.1841$ \\
    \hline
  \end{tabular}
  \label{tab:calcor}
\end{table*}
  
Naturally, one should treat any extrapolation of the correction model into the future, particularly for ACIS-I, with caution.
In addition, as noted above, the correction as defined here may not be appropriate at radii $\gtrsim 3'$ distance from the aimpoint.
Finally, note that the data presented here can only test the correction's efficacy over a limited range in (redshift-corrected) source temperature, $kT/(1+z)\sim4.5$--7.0\,keV, although the good performance over this range provides reason for optimism.
At this writing, the key observation of time dependence in the measured temperatures in the core of A1795 has been verified by the ACIS calibration team (albeit prior to the release of {\sc caldb} 4.12.2; private communication).
Ultimately, we expect this appendix to be obviated by a future update to the Chandra calibration.

\section{Point source sensitivity curves} \label{sec:ptsens}

Table~\ref{tab:ptsens} provides parameters defining the minimum number of counts required for detection with 99 percent probability in a Chandra observation.
For different parts of the focal plane and for a given angular distance from the aimpoint, these curves are quadratic in the logarithm of the background count density (counts per pixel), $\rho$, i.e.
\begin{equation} \label{eq:ptsens}
  \ln \langle N_\mathrm{pt} \rangle = a\ln(\rho)^2 + b\ln(\rho) + c,
\end{equation}%
where in the table each coefficient is subscripted to indicate the CCD numbers it applies to.
Both $\langle N_\mathrm{pt} \rangle$ and $\rho$ refer to averages in the sense of the expectation value of a Poisson distribution.

\begin{table*}
  \caption{Sensitivity Curve Parameters}
    \begin{tabular}{cccccccccccc}
      \hline
      $\theta$ ($'$)  & $a_{01}$ & $b_{01}$ & $c_{01}$ & $a_{23}$ & $b_{23}$ & $c_{23}$ & $a_{567}$ & $b_{567}$ & $c_{567}$\\
      \hline
0 & 0.0208 & 0.338 & 3.73 & 0.0208 & 0.338 & 3.73 & 0.0199 & 0.337 & 3.72\\
3 & 0.0197 & 0.353 & 3.88 & 0.0197 & 0.353 & 3.88 & 0.0210 & 0.368 & 3.99\\
4 & 0.0223 & 0.380 & 4.06 & 0.0223 & 0.380 & 4.06 & 0.0238 & 0.406 & 4.24\\
5 & 0.0251 & 0.422 & 4.29 & 0.0251 & 0.422 & 4.29 & 0.0252 & 0.459 & 4.57\\
6 & 0.0166 & 0.399 & 4.52 & 0.0166 & 0.399 & 4.52 & 0.0194 & 0.451 & 4.79\\
7 & 0.0240 & 0.463 & 4.81 & 0.0240 & 0.463 & 4.81 & 0.0203 & 0.463 & 5.07\\
8 & 0.0153 & 0.447 & 5.00 & 0.0189 & 0.454 & 4.98 & 0.0143 & 0.464 & 5.29\\
9 & 0.0201 & 0.457 & 5.18 & 0.0187 & 0.464 & 5.17 & 0.0218 & 0.503 & 5.50\\
10 & 0.0204 & 0.478 & 5.39 & 0.0143 & 0.457 & 5.38 & 0.0144 & 0.479 & 5.73\\
11 & \ldots & \ldots & \ldots & \ldots & \ldots & \ldots & 0.0181 & 0.488 & 5.90\\
12 & \ldots & \ldots & \ldots & \ldots & \ldots & \ldots & 0.0177 & 0.476 & 6.10\\
13 & \ldots & \ldots & \ldots & \ldots & \ldots & \ldots & 0.0218 & 0.486 & 6.35\\
14 & \ldots & \ldots & \ldots & \ldots & \ldots & \ldots & 0.0201 & 0.473 & 6.60\\
15 & \ldots & \ldots & \ldots & \ldots & \ldots & \ldots & 0.0245 & 0.483 & 6.80\\
      \hline
    \end{tabular}
    \label{tab:ptsens}
\end{table*}



\end{document}